\documentclass[10pt,journal,cspaper,compsoc]{IEEEtranTkde}
\usepackage{graphicx}
\usepackage{caption}
\usepackage{subcaption}
\usepackage{times,amsmath,amsfonts,epsfig, amssymb}
\usepackage{amsmath}
\usepackage[linesnumbered, ruled]{algorithm2e}
\usepackage{url}
\usepackage{xcolor}
\usepackage{balance}
\usepackage{color}
\usepackage{epstopdf}
\usepackage{tabularx}

\newtheorem{example}{\textbf{Example}}
\newtheorem{query}{\textbf{Query}}

\newcommand{\tabincell}[2]{\begin{tabular}{@{}#1@{}}#2\end{tabular}}
\title{Similarity Group-by Operators for Multi-dimensional Relational Data}
\author{Mingjie~Tang, Ruby~Y.~Tahboub,
        Walid~G.~Aref,~\IEEEmembership{Senior Member,~IEEE,} \\
        Mikhail~J.~Atallah,~\IEEEmembership{Fellow,~IEEE and ACM,} Qutaibah~M.~Malluhi,
        Mourad~Ouzzani,~\IEEEmembership{Member,~IEEE,}
        and Yasin~N.~Silva
\IEEEcompsocitemizethanks{\IEEEcompsocthanksitem M. Tang and R. Tahboub is with the Department
of Computer Science, Purdue University, Indiana,
IN, 47906.\protect\\
E-mail: tang49@purdue.edu
\IEEEcompsocthanksitem W.G. Aref is with the Department of Computer Science, Purdue, and Center
for Education and Research in Information Assurance and Security (CERIAS).
\IEEEcompsocthanksitem M. Atallah is with the Department of Computer Science,Purdue University.
\IEEEcompsocthanksitem Q. Malluhi is with the Department of Computer Science and Engineering, Qatar University.
\IEEEcompsocthanksitem M. Ouzzani is with the Qatar Computing Research Institute.
\IEEEcompsocthanksitem Y. Silva is with the School of Mathematical and Natural Science,Arizona State University.}
\thanks{}}
\begin{document}

\IEEEcompsoctitleabstractindextext{
\begin{abstract}

The SQL group-by operator plays an important role in summarizing and aggregating large datasets in a data analytics
stack. While the standard group-by operator, which is based on equality, is useful in several applications, allowing similarity aware
grouping provides a more realistic view on real-world data that could lead to better insights. The Similarity SQL-based
Group-By operator (SGB, for short) extends the semantics of the standard SQL Group-by by grouping data with similar but not
necessarily equal values. While existing similarity-based grouping operators efficiently materialize this “approximate” semantics,
they primarily focus on one-dimensional attributes and treat multi-dimensional attributes independently. However, correlated
attributes, such as in spatial data, are processed independently, and hence, groups in the multi-dimensional space are not
detected properly. To address this problem, we introduce two new SGB operators for multi-dimensional data. The first operator
is the clique (or distance-to-all) SGB, where all the tuples in a group are within some distance from each other. The second
operator is the distance-to-any SGB, where a tuple belongs to a group if the tuple is within some distance from any other tuple
in the group. Since a tuple may satisfy the membership criterion of multiple groups, we introduce three different semantics to
deal with such a case: (i) eliminate the tuple, (ii) put the tuple in any one group, and (iii) create a new group for this tuple. We
implement and test the new SGB operators and their algorithms inside PostgreSQL. The overhead introduced by these operators
proves to be minimal and the execution times are comparable to those of the standard Group-by. The experimental study, based
on TPC-H and a social check-in data, demonstrates that the proposed algorithms can achieve up to three orders of magnitude
enhancement in performance over baseline methods developed to solve the same problem.

\end{abstract}

}

\maketitle

\IEEEdisplaynotcompsoctitleabstractindextext

\IEEEpeerreviewmaketitle

%
\section{Introduction}
\label{introduction}

The deluge of data accumulated from sensors, social networks,
computational sciences, and location-aware services
calls for advanced querying and analytics that are often dependent on
efficient aggregation and summarization techniques.
The SQL group-by operator
is one main construct that is used in
conjunction with aggregate operations to cluster
the data into groups and produce useful summaries.
Grouping is usually performed by aggregating into the same groups tuples
with equal values on a certain subset of the attributes.
However, many applications (i.e.,in Section~\ref{section:applications}) are often interested in grouping based on \textit{similar} rather than strictly equal values.
Clustering~\cite{BIBExample:han2006data}
is a well-known technique for grouping similar data items in the
multi-dimensional space.
In most cases, clustering is performed outside of the database system.
Moving the data outside of the database to perform the clustering
and then back into the database for further processing results in
a costly impedance mismatch.
Moreover, based on the needs of the underlying applications, the output
clusters may need to be further processed by SQL to filter out
some of the clusters and to perform other  SQL operations on the remaining clusters.
Hence, it would be greatly beneficial to develop practical and fast
similarity group-by operators that can be embedded within SQL
to avoid such impedance mismatch and to benefit from the processing power of
all the other SQL operators.

SQL-based Similarity Group-by (SGB) operators have been proposed
in~\cite{BIBExample:silva2013similarity} to support several semantics to group similar but not necessarily equal data.
Although several applications  can benefit from using existing SGB over
Group-by, a key shortcoming of  these operators is that they
focus on one-dimensional data. Consequently, data can only be approximately grouped based on one attribute at a time.





\begin{figure}
\centering
\includegraphics[width=3in,height=1.2in]{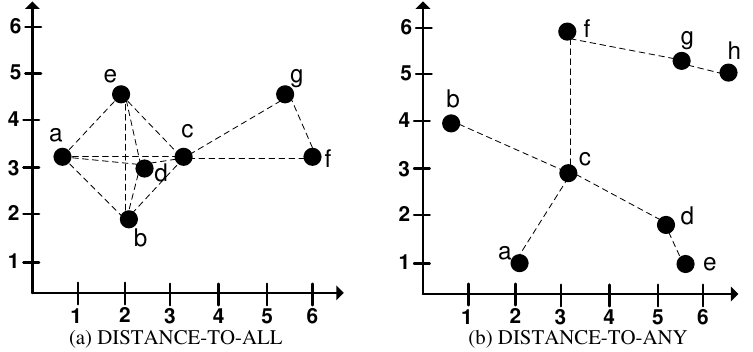}
\caption{The Semantics of Similarity predicates $\epsilon=3$.}
\label{fig:sgbnotion}
\end{figure}

\begin{sloppypar}
In this paper, we introduce new similarity-based group-by operators that group multi-dimensional data using various metric distance functions.
More specifically, we propose two SGB operators, namely SGB-All and
SGB-Any, for grouping multi-dimensional data.
SGB-All forms groups such that a tuple or a data item, say $o$, belongs to a
group, say $g$,
if $o$ is at a distance within a user-defined threshold from all other data items in $g$.
In other words, each group in SGB-All forms a clique of nearby data items
in the multi-dimensional space.
For example, all the two-dimensional points ($a$-$e$)
in Figure~\ref{fig:sgbnotion}a are within distance 3 from each other and hence form a clique.
They are all reported as members of one group as they are all part of the output of SGB-All.
In contrast, SGB-Any forms groups such that a tuple or a data item, say $o$,
belongs to a group, say $g$,
if $o$ is within a user-defined threshold
from at least one other data item in $g$.
For example, all the two dimensional points in Figure~\ref{fig:sgbnotion}b form one group.
Point $a$ is within Distance $3$ from Point $c$ that in turn is within Distance $3$ from Points $b$, $d$, and $f$. Furthermore, Point $e$ is within Distance $3$ from Point $d$, and so on. Therefore, Points a-h of Figure \ref{fig:sgbnotion}b are reported as members of one group as part of the output of SGB-Any.

Notice that in the SGB-All operator,  a data item may qualify the
membership criterion of multiple groups.
For example, data item $c$ in Figure~\ref{fig:sgbnotion}a forms a clique with two groups.
In this case, we propose three semantics, namely,
\textit{on-overlap join-any}, \textit{on-overlap eliminate}, and \textit{on-overlap form-new-group}, for handling such a case.
We provide efficient algorithms for computing the two proposed SGB operators over
correlated multi-dimensional data.
The proposed algorithms use a filter-refine paradigm. In the filter step, a fast yet
conservative check is performed to identify the data items that
are candidates to form groups. Some of the data items resulting
from the filter step will end up being false-positives that will be discarded.
The refinement step eliminates the false-positives to produce the final output groups.
Notice that for the case of SGB-Any, a data item cannot belong to multiple groups.
For example, consider a data item, say $o$, that is a member of two groups, say $g_1$ and $g_2$, i.e., $o$ is within distance $epsilon$ from at least one other data item in each of $g_1$ and $g_2$. In this case, based on the semantics of SGB-Any, Groups $g_1$ and $g_2$ merge into one encompassing bigger group that contains all members of $g_1$, $g_2$ and common data item $o$. Specificity, we mainly focus on two and three dimensional data space, leaving higher dimensions for future work.

The contributions of this paper are  as follows:
\begin{enumerate}

\item We 
	  introduce two
      new operators, namely SGB-All and SGB-Any, for grouping
      multi-dimensional data from within SQL.

\item We present an extensible algorithmic framework to accommodate the
      various semantics of SGB-All and SGB-Any along with
      various options to handle overlapping data  among
      groups. We introduce effective optimizations for both operators.

\item 
	  We prototype the two operators inside PostgreSQL
      and study their performance using the TPC-H benchmark. The
      experiments demonstrate that the proposed algorithms can achieve
up to three orders of magnitude enhancement in performance over the baseline approaches. Moreover, the performance of the proposed SGB operators is comparable to that of relational Group-by, and outperform state-of-the-art clustering algorithm (i.e., \textit{K-means},  \textit{DBSCAN} and \textit{BIRCH}) from one to three orders of magnitude.

\end{enumerate}

\end{sloppypar}

The rest of the paper proceeds as follows.
Section~\ref{section:related-work}
      discusses the related work.
Section~\ref{section:background} provides background
material. Section~\ref{section: similarity operators} introduces the
new SGB operators.
Section~\ref{section:applications} presents application scenarios that demonstrate the use and practicality of the various proposed semantics for SGB operators.
Sections~\ref{section:sgball-framework} and ~\ref{section:sgb-any-framework}
introduce the algorithmic frameworks for SGB-All and SGB-Any operators, respectively.
Section~\ref{section:performance-evaluation}
describes the in-database extensions to support the
two operators and their performance evaluation from within PostgreSQL.
Section~\ref{section:conclusion} concludes the paper.

\section{Related Work}
\label{section:related-work}

\begin{sloppypar}
Previous work on similarity-aware query processing addressed the
      theoretical foundation and query optimization issues for
      similarity-aware query operators~\cite{BIBExample:silva2013similarity}.
      \cite{BIBExample:adali1998multi,BIBExample:atnafu2001similarity}
      introduce similarity algebra that extends relational algebra
      operations, e.g., joins and set operations, with similarity
      semantics. Similarity queries and their optimizations include
      algorithms for similarity range search and K-Nearest Neighbor
      (KNN)~\cite{BIBExample:braunmuller2001multiple}, similarity
      join~\cite{BIBExample:chen2003similar_join}, and similarity
      aggregates~\cite{BIBExample:razente2008aggregate}.
Most of work focus on semantic and transformation rules for
query optimization purpose independently from actual algorithms to realize similarity-aware operators.
In contrast, our focus is on the latter.
\end{sloppypar}

Clustering forms groups of similar data for the purpose of learning hidden
      knowledge.
Clustering methods and algorithms have been extensively studied in the
      literature, e.g.,
see~\cite{BIBExample:berkhin2006survey,BIBExample:han2006data}.
The main clustering methods are
partitioning,
hierarchical, and
density-based.
\textit{K-means}~\cite{BIBExample:kanungo2002efficient}
is a widely used partitioning algorithm that uses several
      iterations to refine the output clusters.
Hierarchical methods build clusters either divisively (i.e., top-down) such as in
      \textit{BIRCH}~\cite{BIBExample:zhang1996birch},
or agglomeratively (i.e., bottom-up) such  as in \textit{CURE}~\cite{BIBExample:guha1998cure}.
Density-based methods, e.g., \textit{DBSCAN}~\cite{BIBExample:ester1996density}, cluster
      data based on local criteria, e.g., density reachability among
      data elements.
The key differences between our proposed SGB operators and clustering are:
(1)~the proposed SGB operators are relational  operator that are integrated in
a relational query evaluation pipeline with various grouping semantics.
Hence, they avoid the impedance mismatch
experienced by standalone clustering and data mining packages that mandate
extracting the data to be clustered out of the DBMS.
(2)~In contrast to standalone clustering algorithms, the SGB operators can be
      interleaved with other
relational operators. 
(3)~Standard relational query optimization techniques that apply to the standard relational group-by are also
      applicable
to the SGB operators as illustrated in~\cite{BIBExample:silva2013similarity}.
This is not feasible with
standalone clustering algorithms. Also, improved performance
can be gained by using database access methods that process multi-dimensional data.

An early work       
on similarity-based
      grouping appears
in~\cite{BIBExample:schallehn2004efficient}.
It  addresses the inconsistencies and redundancy encountered while
      integrating information systems with dirty data.
However, this work 
realizes similarity grouping through  pairwise comparisons
which incur excessive computations in the absence of
      a proper index. Furthermore, the introduced extensions
      are not integrated as first class database operators.
The work
      in~\cite{BIBExample:zhang2007cluster}
%
focuses on overcoming the limitations of the distinct-value group-by
      operator and introduces
    the SQL
      construct ``Cluster~By" that uses conventional clustering
      algorithms, e.g., \textit{DBSCAN}, to realize similarity grouping.
Cluster~By addresses the impedance mismatch  due to
      the data being outside  the DBMS to perform clustering.
Our SGB operators 
      are more generic as they use a set of predicates and clauses to refine
      the grouping semantics, e.g., the distance relationships among the data elements
      that constitute the group and how inter-group overlaps are dealt with.

\begin{sloppypar}
Several DBMSs have been extended to support 		similarity operations.
SIREN~\cite{BIBExample:razente2006siren} is a 	   similarity retrieval engine that allows executing similarity queries 	 over a relational DBMS. 
POSTGRESQL-IE~\cite{BIBExample:guliato2009postgresql} is an image 	  handling extension of PostgreSQL to support content-based image 		
retrieval capabilities, e.g., supporting the image data type and 		  responding to image similarity queries.
While these extensions
incorporate various notions of similarity into query processing,
they focus on the 		  similarity search operation.
SimDB~\cite{BIBExample:silva2013similarity} is a PostgreSQL extension that 		
supports similarity-based queries and their optimizations.
Several similarity operations, e.g., join and group-by, are implemented in as first-class database operators.
However, the similarity operators in SimDB focus on one-dimensional data and do not handle multi-dimensional attributes.

\end{sloppypar}

\section{Preliminaries}
\label{section:background}

In this section, we provide background definitions and formally introduce similarity-based group-by operators.

\newtheorem{defn}{Definition}

\begin{defn}
A \textbf{metric space} is a space $\textbf{M} = \langle \mathbb{D}, \delta\rangle$ in which the distance between
two data points within a domain $\mathbb{D}$ is defined by a function $\delta : \mathbb{D} \times \mathbb{D} \to
\mathbb{R}$ that satisfies the properties of symmetry, non-negativity, and triangular inequality.
\end{defn}

We use the Minkowski distance $L_p$ as the distance function $\delta$.
We consider the following two Minkowski distance functions.
Let $p_x$ be a data point in the multi-dimensional space of the form $p_x:\langle x_1, ..., x_d\rangle$ and $p_{xy}$ is the value of the $y^{th}$ dimension of $p_x$.
\begin{itemize}
\item The Euclidean distance\\ $L_2 : \delta_2(p_i, p_j) = \sqrt{\displaystyle\sum_y\left(p_{iy}-p_{jy}\right)^2}$
\item The maximum distance \\$L_\infty : \delta_\infty(p_i, p_j) =
\displaystyle \max_y\left|p_{iy}-p_{jy}\right|$.
\end{itemize}

\begin{defn}
A \textbf{similarity predicate} $\xi_{\delta, \epsilon}$ is a Boolean expression that returns TRUE for two multi-dimensional points, say $p_i$ and $p_j$, if
the distance $\delta$ between $p_i$ and $p_j$ is less than or equal to $\epsilon$,
i.e.,
$
	\xi_{\delta,\epsilon}(p_i, p_j) : \delta(p_i, p_j) \le \epsilon.
$
~In this case, the two points are said to be similar.
\end{defn}

\begin{sloppypar}
\begin{defn}
Let T be a relation of tuples, where each tuple, say $t$, is of the form
$t = \left\lbrace GA_{1}, ..., GA_{k},NGA_{1}, ..., NGA_{l}, \right\rbrace $,
 the subset
$
GA_c = \left\lbrace GA_1, ..., GA_k \right\rbrace
$ be the grouping attributes, the subset
$
NGA = \left\lbrace NGA_1, ..., NGA_l \right\rbrace
$ be the non-grouping attributes, and
${\xi}_{\delta,\epsilon}$ be a similarity predicate. Then, the \textbf{similarity Group-by operator}
$
{\mathcal{G}}_{\langle GA_c, ({\xi}_{\delta, \epsilon})\rangle} (R)
$
forms a set of answer groups $G_{s}$ by applying
${\xi}_{\delta, \epsilon}$ to the elements of $GA_c$ such that a pair of tuples, say $t_i$ and $t_j$, are in the same group if $ \xi_{\delta, \epsilon}(t_i._{GAc}, t_j._{GAc})$.
\end{defn}
\end{sloppypar}
\begin{sloppypar}

\begin{defn}
\label{def:oset}
Given a set of groups $G=\{g_1, ..., g_m\}$,
the \textbf{Overlap Set} $Oset$ is the set of tuples formed by the union of the intersections of all pairs of groups $(g_1, ..., g_m)$,
i.e., 
$Oset =\cup_{(i,j) \in \{1..m\}} (g_i \cap g_j )$, where $i \neq j$. In other words, $Oset$ contains all the tuples that belong to more than one group.
\end{defn}
\end{sloppypar}

For simplicity, we study the case where the set of grouping attributes, $GA_c$, contains only two attributes.
In this case, we can view tuples as points in the two-dimensional space, each of the form $p$:$(x_1,x_2)$.
We enclose each group of points by a bounding rectangle $R$:$(p_l, p_r)$, where
points $p_l$ and $p_r$ correspond to the upper-left and bottom-right corners of R, respectively.

\section{Similarity Group-By Operators}
\label{section: similarity operators}

This section introduces the semantics of the two similarity-based group-by operators, namely, SGB-All and SGB-Any.

\subsection{Similarity Group-By ALL (SGB-All)}

\begin{sloppypar}
Given a set of tuples whose grouping attributes form a set, say $P$, of two-dimensional points, where $P=\left\lbrace p_1, ..., p_n \right\rbrace$,
the SGB-All operator $\mathcal{\check{G}}_{all}$ forms a set, say $G_m$, of groups  of points from $P$ such that
$\forall g\in G_m$, the similarity predicate $\xi _{\delta, \epsilon}$ is TRUE for all pairs of points $\langle p_i ,p_j\rangle \in g$, and $g$ is maximal, i.e, there is no group $g'$ such that $g \subseteq g'$.
More formally,
\[\mathcal{\check{G}}_{all} =\left\lbrace g \;| \; \forall p_{i}, p_{j} \in g, \; \xi _{\delta, \epsilon}(p_i, p_j) \wedge g \; is \;maximal\right\rbrace\]
\end{sloppypar}

Figure~\ref{fig:sgbnotion} gives an example of two groups (a-e) and (c,f,g), where all pairs of elements within each group are within a distance $\epsilon \le 3$. The proposed SQL syntax for the SGB-All operator is as follows:\\
{\ttfamily \scriptsize
\hspace*{4ex}SELECT \textit{column}, aggregate-func(\textit{column})
\newline
\hspace*{4ex}FROM \textit{table-name}
\newline
\hspace*{4ex}WHERE \textit{condition}
\newline
\hspace*{4ex}{GROUP BY}
\textit{column} \textbf{DISTANCE-TO-ALL} [\textit{L2} $ \mid $ \textit{LINF}] \textbf{WITHIN} \textit{$\epsilon$}
\newline
\hspace*{4ex}\textbf{ON-OVERLAP} [\textit{JOIN-ANY} $ \mid $ \textit{ELIMINATE} $ \mid $\textit{FORM-NEW-GROUP}]
}

SGB-All uses the following clauses to realize similarity-based grouping:
\begin{itemize}
\item
DISTANCE-TO-ALL: specifies the
distance function to be applied by the similarity predicate when deciding the membership of points within a group.
\begin{itemize}
\item L2: $L_2$ (Euclidean distance).
\item LINF: $L_{\infty}$ (Maximum infinity distance)
\end{itemize}

\item ON-OVERLAP: is an arbitration clause to decide on a course of action when
a data point is within Distance $\epsilon$ from more than one group.
When a point, say $p_i$, matches the membership criterion for more than one group, say $g_1 \cdots g_w$,
one of the three following actions are taken:
\begin{itemize}

\item JOIN-ANY: the data point $p_i$ is randomly inserted into any one group out of $g_1 \cdots g_w$.

\item ELIMINATE: discard the  data point $p_i$, i.e.,
all data points in $Oset$ (see Definition \ref{def:oset}) are eliminated.

\item FORM-NEW-GROUP: insert $p_i$ into a separate group, i.e.,
form new groups out of the points in $Oset$.

\end{itemize}
\end{itemize}

\begin{figure}
\centering
\includegraphics[width=1.6in,height=1.25in]{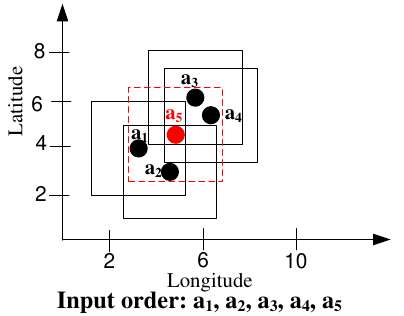}
\caption{Data points using $\epsilon = 3$ and $L_{\infty}$.}
\label{fig:sgbexample}
\end{figure}

\begin{sloppypar}
\begin{example}
\label{example-all} 
The following query  performs the aggregate operation $count$ on the groups formed by SGB-All
on the two-dimensional grouping attributes \textit{GPSCoor-lat} and \textit{GPSCoor-long}.
The $L_{\infty}$ distance is used
with Threshold $\epsilon=3$.\\
{\ttfamily \scriptsize
\hspace*{4ex}SELECT \textit{count(*)}
\newline
\hspace*{4ex}FROM \textit{$GPSPoints$}
\newline
\hspace*{4ex}GROUP BY \textit{GPSCoor-lat,GPSCoor-long} \textbf{DISTANCE-TO-ALL} \textit{LINF}
\newline
\hspace*{4ex}\textbf{WITHIN} \textit{3}
\newline
\hspace*{4ex}\textbf{ON-OVERLAP} <\textit{clause}>
}

Consider Points $a_1$-$a_5$ in Figure~\ref{fig:sgbexample} that arrive in the order $a_1,a_2, \cdots,a_5$.
After processing $a_4$, the following groups satisfy the SGB-All predicates: $g_{1} \left\lbrace a_{1}, a_2 \right\rbrace$ and $g_{2} \left\lbrace a_{3},a_{4} \right\rbrace$. However, Data-point $a_5$ is within $\epsilon$ from $a_1,a_2$ in $g_1$ and $a_3, a_5$ in $g_2$. Consequently, an arbitration ON-OVERLAP clause is necessary. We consider the three possible semantics below for illustration.

With an ON-OVERLAP JOIN-ANY clause,
a group is selected at random. If $g_1$ is selected,
the resulting groups are $g_1\{a_1, a_2, a_5\}$ and $g_2\{a_3, a_4\}$,
and the answer to the query is $\{3,2\}$.
With an ON-OVERLAP ELIMINATE clause, the overlapping point $a_5$ gets dropped; the resulting groups are $g_{1} \left\lbrace a_{1}, a_2 \right\rbrace$ and $g_{2} \left\lbrace a_{3},a_{4} \right\rbrace$, and the query output is $\{2,2\}$.
With an ON-OVERLAP FORM-NEW-GROUP clause, the overlapping point $a_5$ is inserted into a newly created group; the resulting groups are $g_{1} \left\lbrace a_{1}, a_2 \right\rbrace$, $g_{2} \left\lbrace a_{3},a_{4} \right\rbrace$, $g_3\{a_5\}$ and the query output is $\{2,2,1\}$.
\end{example}
\end{sloppypar}

\subsection{Similarity Group-By Any (SGB-Any)}
\begin{sloppypar}

Given a set of tuples whose grouping attributes from a set, say $P$, of two dimensinal points, where $P=\left\lbrace p_1, ..., p_n \right\rbrace$,
the SGB-Any operator $\mathcal{\check{G}}_{any}$ clusters points in $P$ into a set of groups, say $G_m$, such that,
for each group $g \in G_m$, the points in $g$ are all connected by edges to form a graph, where an edge connects two points, say $p_i$ and $p_j$, in the graph if they are within Distance $\epsilon$ from each other, i.e,. $\xi _{\delta, \epsilon}(p_i, p_j)$. More formally,\\

$\mathcal{\check{G}}_{any} =\{ g \;| \; \forall p_{i}, p_{j} \in g,( \xi _{\delta, \epsilon}(p_i, p_j) \; \vee \;(\exists \; p_{k1},..., p_{kn}, \; \xi _{\delta \epsilon} (p_i, p_{k1})\;\wedge ... \wedge \xi _{\delta \epsilon}(p_{kn}, p_{j})) ) \wedge g \; is \;maximal\}$\\

\end{sloppypar}

The notion of distance-to-any between elements within a group is illustrated in Figure \ref{fig:sgbnotion}b, where $\epsilon = 3$. All of the points (a-h) form one group. The corresponding SQL syntax of the SGB-Any operator is as follows:\\
{\ttfamily \scriptsize
\hspace*{4ex}SELECT \textit{column}, aggregate-func(\textit{column})
\newline
\hspace*{4ex}FROM \textit{table-name}
\newline
\hspace*{4ex}WHERE \textit{condition}
\newline
\hspace*{4ex}GROUP BY \textit{column} \textbf{DISTANCE-TO-ANY} [\textit{L2} $\mid$
\textit{LINF}] \textbf{WITHIN} \textit{$\epsilon$} \\ }
SGB-Any uses the DISTANCE-TO-ANY predicate that applies the metric space function while evaluating the distance between adjacent points. When using the semantics for SGB-Any, the case for points overlapping multiple groups does not arise. The reason is that when an input point overlaps multiple groups, the
groups merge to form one large group.

\begin{sloppypar}
\begin{example}
\label{example-any} 
The following query performs the aggregate operation $count$ on the groups formed by SGB-Any on the two-dimensional grouping attributes \textit{GPSCoor-lat} and \textit{GPSCoor-long} using the Euclidean distance with $\epsilon=3$.\\
{\ttfamily \scriptsize
\hspace*{4ex}SELECT \textit{count(*)}
\newline
\hspace*{4ex}FROM \textit{$GPSPoints$}
\newline
\hspace*{4ex}\textbf{GROUP BY} \textit{\textit{GPSCoor-lat} and \textit{GPSCoor-long}} \
\newline
\hspace*{4ex}\textbf{DISTANCE-TO-ANY }\textit{L2} \textbf{WITHIN} \textit{3}
}

Consider the example in Figure~\ref{fig:sgbexample}.
After processing $a_4$, the following groups are $g_1\{ a_1, a_2 \}$ and $g_2\{a_3, a_4\}$.
Since Point $a_5$ is within $\epsilon$ from both $a_1, a_2$ in $g_1$ and $a_3, a_4$ in $g_2$,
the two groups are merged into a single group.
Therefore, the output of the query is $\{5\}$.
Any overlapping point will cause groups to merge and hence there is no need to add a special clause to handle overlaps.
\end{example}
\end{sloppypar}

\section{Applications}
\label{section:applications}

\begin{sloppypar}
In this section, we present application scenarios that demonstrate the practicality and the use of the various semantics for the proposed Similarity Group-by operators.

\begin{figure}
\centering
\includegraphics[width=3.3in,height=1.15in]{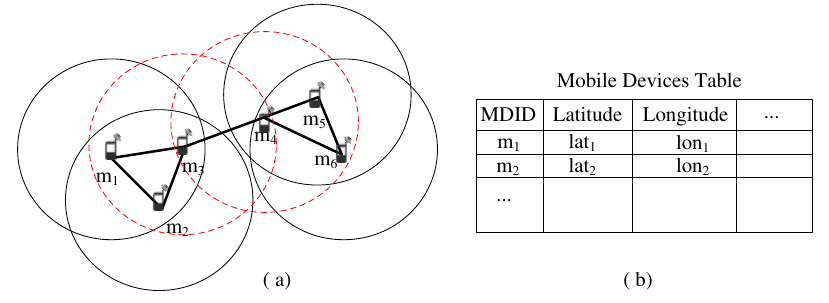}
\caption{(a)~An Mobile Ad hoc Network (MANET) containing the devices $m_1\dots m_6$, where the circle around each device is its signal range, (b)~The corresponding Mobile Devices table.}
\label{fig:manet}
\end{figure}

\begin{example}
\label{example-manet} 
\textbf{Mobile Ad hoc Network (MANET)} is a self-configuring wireless network of mobile devices (e.g., personal digital assistants). A mobile device in a MANET communicates directly with other devices that are within the range of the device's radio signal or indirectly with distant mobile devices using gateways (i.e., intermediate mobile devices, e.g., $m_1$ and $m_2$ in Figure \ref{fig:manet}a). In a MANET, wireless links among nearby devices are established by broadcasting special messages. Radio signals are likely to overlap. As a result, uncareful broadcasting may result in redundant messages, contention, and collision on communication channels 
Consider the Mobile Devices table in Figure~\ref{fig:manet}b that maintains the geographic locations of the mobile devices in a MANET. In the following, we give example queries that illustrate how MANETs can tremendously benefit from SGB-All and SGB-Any operators.
\end{example}

\begin{query}
\label{query:manet-any}
\textbf{Geographic areas that encompass a MANET.} A mobile device, say $m$, belongs to a MANET if and only if $m$ is within the \textit{signal range} from at least one other device mobile. The SGB-ANY semantics identifies a connected group of mobile devices using \textit{signal range} as a similarity grouping threshold.\\
{\ttfamily \scriptsize
\hspace*{4ex}SELECT \textit{ST\_Polygon(Device-lat, Device-long)}
\newline
\hspace*{4ex}FROM \textit{$MobileDevices$}
\newline
\hspace*{4ex}\textbf{GROUP BY} \textit{\textit{Device-lat}, \textit{Device-long}}
\newline
\hspace*{4ex}\textbf{DISTANCE-TO-ANY }\textit{L2} \textbf{WITHIN} \textit{SignalRange}
}
\\
Referring to the mobile devices in Figure \ref{fig:manet}a, the output of Query~\ref{query:manet-any} returns a polygon that encompasses mobile devices $m_1$-$m_6$.
\end{query}

\begin{query}
\label{query:gateway}
\textbf{Candidate gateway mobile devices.} A gateway represents an overlapping mobile device that connects two devices that are not within each other's signal range. The SGB-All FORM-NEW-GROUP inserts the overlapped devices into a new group. Therefore, those devices in the newly formed group are ideal gateway candidates. 
\\
{\ttfamily \scriptsize
\hspace*{4ex}SELECT \textit{COUNT(*)}
\newline
\hspace*{4ex}FROM \textit{$MobileDevices$}
\newline
\hspace*{4ex}\textbf{GROUP BY} \textit{\textit{Device-lat} , \textit{Device-long}}
\newline
\hspace*{4ex}\textbf{DISTANCE-TO-ALL }\textit{L2} \textbf{WITHIN} \textit{SignalRange}
\newline
\hspace*{4ex}\textbf{ON-OVERLAP FORM-NEW-GROUP}
}
\\The output of Query~\ref{query:gateway} returns the number of candidate gateway mobile devices. Along the same line, identifying mobile devices that cannot serve as a gateway is equally important to a MANET. SGB-All ELIMINATE identifies mobile devices that cannot serve as a gateway by discarding the overlapping mobile devices.
\end{query}

\begin{example}
\label{example-lSNE} 
\textbf{Location-based group recommendation in mobile social media}.
Several social mobile applications, e.g., \textit{WhatsApp} and \textit{Line}, employ the frequent geographical location of users to form groups that members may like to join. For instance, users who reside in a common area (e.g., within a distance threshold) may share similar interests and are inclined to share news.
However, members who overlap several groups may disclose information from one group to another and undermine the privacy of the overlapping groups. Query~\ref{query:social-sgball} demonstrates how SGB-ALL allows forming location-based groups without compromising privacy.
\end{example}

\begin{query}
\label{query:social-sgball} \textbf{Forming private location-based groups}. The various SGB-All semantics form groups while handling ON-OVERLAP options that restrict members to join multiple groups. In  Query \ref{query:social-sgball}, we assume that Table Users-Frequent-Location maintains the users' data, e.g., user-id and frequent location. The user-defined aggregate function \textit{List-ID} returns a list that contains all the user-ids within a group.
\\
{\ttfamily \scriptsize
\hspace*{4ex}SELECT \textit{List-ID(user-id)},
\newline
\hspace*{4ex}\textit{ST\_Polygon(User-lat, User-long)}
\newline
\hspace*{4ex}FROM \textit{$Users-Frequent-Location$}
\newline
\hspace*{4ex}\textbf{GROUP BY} \textit{\textit{User-lat} , \textit{User-long}}
\newline
\hspace*{4ex}\textbf{DISTANCE-TO-ALL }\textit{L2} \textbf{WITHIN} \textit{Threshold}
\newline
\hspace*{4ex}\textbf{[ON-OVERLAP JOIN-ANY | ELIMINATE | FORM-NEW-GROUP]}
}\\
The output of Query~\ref{query:social-sgball} returns a list of user-ids for each formed group along with a polygon that encompasses the group's geographical location. The JOIN-ANY semantics recommends any one arbitrary group for overlapping members who in this case will not be able to join multiple groups. The ELIMINATE semantics drops overlapping members from recommendation, while FORM-NEW-GROUP creates dedicated groups for overlapping members.
\end{query}

\end{sloppypar}

\section{Algorithms for SGB-All}
\label{section:sgball-framework}
In this section, we present an extensible algorithmic framework to realize similarity-based grouping using the distance-to-all semantics with the various options to handle the overlapping data among the groups.

\subsection{Framework}

\begin{sloppypar}
Procedure~\ref{alg:sgball-framework} illustrates a generic algorithm to realize SGB-All.
This generic algorithm supports the various data overlap semantics using one algorithmic framework. The algorithm
breaks down
the SGB-All operator into procedures that can be optimized independently.
For each data point, the algorithm starts by identifying two sets (Line 2).
The first set, namely $CandidateGroups$, consists of groups that $p_i$ can join.
$p_i$ can join a group, say $g$, in $CandidateGroups$ if the similarity predicate is true for all pairs $\langle p_i, p_i'\rangle~\forall p_i' \in g$.
The second set, namely $OverlapGroups$, includes groups that have some (but not all) of its data points satisfying the similarity predicate.
A group, say $g$, is in $OverlapGroups$ if there exist at least two points $p$ and $q$ in $g$ such that the similarity distance between $p_i$ and $p$ holds and the similarity distance between $p_i$ and $q$ does not hold.
$OverlapGroups$ serves as a preprocessing step required to handle the semantics of ELIMINATE and FORM-NEW-GROUP encountered in later steps.
Figure~\ref{fig:sgbpoints} gives four existing groups $g_1$-$g_4$ while Data-point $x$ is being processed.
In this case, $CandidateGroups$ contains $\left\lbrace g_2, g_3 \right\rbrace$ and $OverlapGroups$ contains $\left\lbrace g_1 \right\rbrace$.

Procedure $ProcessGroupingALL$ (Line 3 of Procedure \ref{alg:sgball-framework}) uses $CandidateGroups$ and the ON-OVERLAP clause $CLS$ to either
(i)~place $p_i$  into a new group,
(ii)~place $p_i$ into existing group(s),
or (iii)~drop $p_i$ from the output, in case of an ON-OVERLAP
clause.
Finally, Procedure $ProcessOverlap$ (Line 5) uses $OverlapGroups$ to verify whether additional processing is needed to fulfill the semantics of SGB-All.

\end{sloppypar}

\begin{algorithm}[t]\small
\KwIn{$P$: set of data points, $\epsilon$: similarity threshold , $\delta$: distance function , $CLS$: ON-OVERLAP clause, $G$ set of existing groups}
\KwOut{Set of output groups }
\For{ each data element $p_{i}$ in $P$}
{
 $(CandidateGroups, OverlapGroups) \leftarrow FindCloseGroups(p_i, G, \epsilon, \delta, CLS)$\\
 $ProcessGroupingALL(p_i, CandidateGroups, CLS)$\\
\If{CLS is not JOIN-ANY And sizeOf(OverlapGroups)!= 0}
 {$ProcessOverlap(p_i, OverlapGroups, CLS)$}
}
\caption{\small{Similarity Group-By ALL Framework}}
\label{alg:sgball-framework}
\end{algorithm}

\subsection{Finding Candidate and Overlap Groups}
\label{section: Finding Candidate and Overlap Groups }
\begin{sloppypar}
In this section, we present a straightforward approach to identify $CandidateGroups$ and $OverlapGroups$. In Section~\ref{subsection:boundschecking}, we propose a new two-phase filter-refine approach that utilizes a conservative check in the filter phase to efficiently identify the member groups in $CandidateGroups$. Then, in Section ~\ref{section:false-positive}, we introduce the refine phase that is applied only if $L_2$ is used as the distance metric to detect the $CandidateGroups$ that falsely pass the filter step.

Procedure~\ref{alg:findcloseall-nested} gives the pseudocode for \textit{Naive FindCloseGroups}
that evaluates the distance-to-all similarity predicate between $p_i$ and all the points that have been previously processed (Lines 6-15).
The grouping semantics (Lines 16-20) identify how the two sets $CandidateGroups$ and $OverlapGroups$ are populated.
\end{sloppypar}

\begin{algorithm}[t]\small
\KwIn{$p_i$: data point, $\epsilon$: similarity threshold , $\delta$: distance function, $CLS$: ON-OVERLAP clause, $G$: set of existing groups }
\KwOut{$Candidate$, $OverlapGroups$}
$Candidate\leftarrow NULL$\\
$OverlapGroups\leftarrow NULL$\\
\For{ each group $g_{j}$ in $G$}
{
CandidateFlag = True\\
OverlapFlag = False\\
\For{ each $p_{k}$ in $g_j$}
{\uIf { (Distance($p_i$, $p_k$, $\delta$)$\leqslant \epsilon$)}{OverlapFlag = True
}\Else {CandidateFlag =  False \\ \If{CLS == JOIN-ANY}{break}}

}
\uIf{CandidateFlag is True}{ insert $g_j$ into $Candidate$}
\ElseIf {CLS is not JOIN-ANY and CandidateFlag is False and OverlapFlag is True}{insert $g_j$ into $OverlapGroups$}
}
\caption{\small{Naive FindCloseGroupsALL}}
\label{alg:findcloseall-nested}
\end{algorithm}

\subsubsection{Processing New Points}
The second step of the SGB-All Algorithm in Procedure \ref{alg:sgball-framework} places $p_i$, the data point being processed, into a new group or into an existing group, or drops $p_i$ from the output depending on the semantics of SGB-All specified in the query.

\begin{figure}[t]
\centering
\includegraphics[width=2.8in,height=1.3in]{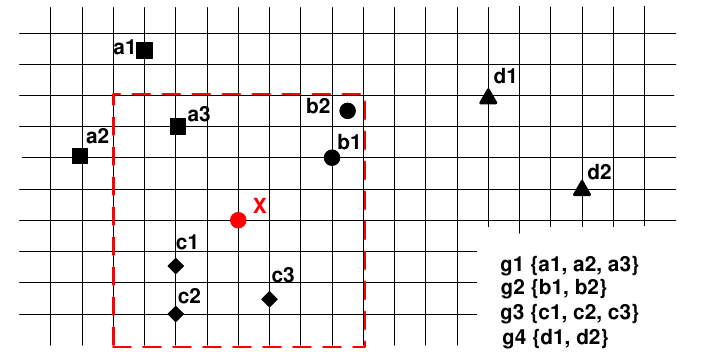}
\caption{Processing the point x using $L_{\infty}$ with $\epsilon = 4$. }
\label{fig:sgbpoints}
\end{figure}

\begin{sloppypar}
Procedure~\ref{alg:processgroupingall} (ProcessGroupingAll) proceeds as follows. First, it identifies the cases where $CandidateGroups$ is empty or consists of a single group. In these cases, $p_i$ is inserted into a newly created group or into an existing group depending on $p$'s distance from the existing group. Procedure $ProcessInsert$ places the data point $p_i$ into a group.
Next, the ON-OVERLAP clause $CLS$ is consulted to determine the proper course of action. The JOIN-ANY clause arbitrates among the overlapping groups by inserting $p_i$ into a randomly chosen group. The procedure $ProcessEliminate$ (Line 13) handles the details of processing the ELIMINATE clause.
Consider the example illustrated in Figure \ref{fig:sgbpoints}, where $CandidateGroups$ consists of $ \left\lbrace g_2, g_3 \right\rbrace$. $ProcessEliminate$ drops Point $x$.

Finally, Procedure $ProcessNewGroup$ (Line 15) processes the FORM-NEW-GROUP clause. It inserts $p_i$ into a temporary set termed $S'$ for further processing. The SGB-All with FORM-NEW-GROUP option forms groups out of $S'$ by calling SGB-All recursively until $S'$ is empty.

\end{sloppypar}

\begin{algorithm}[t]\small
\KwIn{$p_i$: data point, $CLS$: ON-OVERLAP clause, $CandidateGroups$ }
\KwOut{updates $CandidateGroups$ based on $CLS$ semantics}
\uIf{sizeof(CandidateGroups)== 0 }{ create a new group $g_{new}$ \\
 $ProcessInsert(p_i, g_{new})$}
\uElseIf {sizeof(CandidateGroups) == 1 }{ insert into existing group $g_{out}$\\  $ProcessInsert(p_i, g_{out})$ }
\Else{\Switch{CLS}{
\uCase{JOIN-ANY}{
 $g_{out} \leftarrow GetRandomGroup(CandidateGroups)$\\
 $ProcessInsert(p_i, g_{out})$
}
\uCase{ELIMINATE}{
ProcessEliminate($p_i$, $CandidateGroups$)
}
\uCase{FORM-NEW-GROUP}{
ProcessNewGroup($p_i$, $CandidateGroups$)
}
}}

\caption{\small{ProcessGroupingALL}}
\label{alg:processgroupingall}
\end{algorithm}

\subsubsection{Handling Overlapped Points}
\begin{sloppypar}
The final step of SGB-All in Procedure~\ref{alg:sgball-framework} processes the groups in the Set $OverlapGroups$.  $OverlapGroups$ consists of groups, where each group has some data points (but not all of them) that satisfy the similarity predicate with the new input point $p_i$. This step is required by the ELIMINATE and FORM-NEW-GROUP semantics.
Procedure $ProcessOverlap$ handles the ELIMINATE semantics as follows. It iterates over $OverlapGroups$ and deletes overlapped data points.
Consider the example illustrated in Figure \ref{fig:sgbpoints}. Set $OverlapGroups$ consists of $ \left\lbrace g_1 \right\rbrace$ with overlapped Data-Point $a_3$.
Finally, $ProcessOverlap$ handles the FORM-NEW-GROUP semantics by inserting the overlapped data points into a temporary set termed $S'$ and deletes these points from their original groups.
\end{sloppypar}

\begin{sloppypar}
 The time complexity for SGB-All according the algorithmic framework in Procedure \ref{alg:sgball-framework} is dominated by the time complexity of $FindCloseGroups$. The time complexity of $ProcessGrouping$ and $ProcessOverlap$ (Lines 3-6) is linear in the size of $CandidateGroups$  and $OverlapGroups$. Consequently, given an input set of size $n$, Procedure \textit{Naive FindCloseGroups} incurs $n \choose 2$ distance computations that makes the upper-bound time complexity of SGB-All quadratic i.e., $O(n^2)$.
Section \ref{subsection:boundschecking} introduces a filter-refine paradigm to optimize over Procedure \textit{Naive FindCloseGroups}.
\end{sloppypar}

\subsection{The Bounds-Checking Approach}
\label{subsection:boundschecking}

\begin{figure}[t]
\centering
\includegraphics[width=3.26in,height=2.15in]{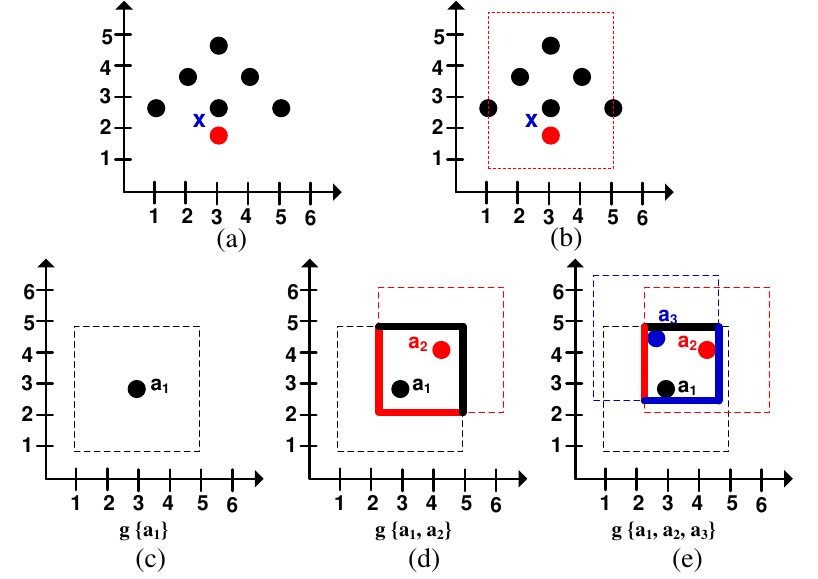}
\caption{The $\epsilon$-All Bounding Rectangle approach.}
\label{fig:bounds-check-new}
\end{figure}

In this section, we introduce a  Bounds-Checking approach to optimize over Procedure
\textit{Naive FindCloseGroups}. Consider the data points of Group $g$ illustrated in Figure ~\ref{fig:bounds-check-new}a.
Procedure~\textit{Naive FindCloseGroups} performs six distance computations to determine whether a new data point $x$ can join Group $g$. To reduce the number of comparisons, we introduce a bounding rectangle for each Group $g$ in conjunction with the similarity threshold $\epsilon$ so that all data points that are bounded by the rectangle satisfy the distance-to-all similarity predicate. For example, Data Element $x$ in Figure \ref{fig:bounds-check-new}b is located inside $g$'s bounding rectangle. Therefore, $g$ is a candidate group for $x$.

\begin{defn}
Given a set of multi-dimensional points and a similarity predicate $\xi_{\delta_\infty, \epsilon}$, the \textbf{$\epsilon$-All Bounding Rectangle} $R_{\epsilon-All}$ is a bounding rectangle such that for any two points $x_i$ and $y_i$ bounded by $R_{\epsilon-All}$, the simiarity predicate $\xi_{\delta_{\infty},\epsilon}(x_i, y_i)$ is true.
\end{defn}

Consider Figure~\ref{fig:bounds-check-new}c, where the bounding rectangle $R_{\epsilon-All}$ is constructed for a group that consists of a single Point $a_1$, where $\epsilon = 2$ and the sides of the rectangle are $2\epsilon$ by $2\epsilon$ centered at $a_1$. After inserting the second Point $a_2$ into $g$, as in Figure~\ref{fig:bounds-check-new}d,
$R_{\epsilon-All}$ is shrunk to include the area where the similarity predicate is true for both Points $a_1$ and $a_2$. The invariant that $R_{\epsilon-All}$ maintains varies depending on the distance metric used.
For the $L_{\infty}$ distance metric, $R_{\epsilon-All}$ is updated such that if a Point, say $x_i$, is inside $R_{\epsilon-All}$, then $x_i$ is guaranteed to be within Distance $\epsilon$ from all the points that form Group $g$. For the Euclidean distance,
the invariant that $R_{\epsilon-All}$ maintains is that if a point, says $x_i$, is outside $R_{\epsilon-All}$, then $x_i$ cannot belong to Group $g$. In this case, if $x_i$ is inside $R_{\epsilon-All}$, it is likely that
$x_i$ is within distance $\epsilon$ from all the points inside $R_{\epsilon-All}$. Hence, for the Euclidean distance, $R_{\epsilon-All}$ is a conservative representation of the group $g$ and serves as a filter step to save needless comparisons for points that end up being outside of the group.
We illustrate in Figures~\ref{fig:bounds-check-new}c-~\ref{fig:bounds-check-new}e how to maintain these invariants when a new point joins the group. We use the case of $L_{\infty}$ for illustration.
When a new point $x_i$ is inside the bounding rectange $R_{\epsilon-All}$ of Group $g$, then $x_i$ is within Distance $\epsilon$ from all the points in the group, and hence will join Group $g$. Once $x_i$ joins Group $g$, the bounds of Rectangle $R_{\epsilon-All}$ are updated to retain the truth of $R_{\epsilon-All}$'s invariant.
The sides of $R_{\epsilon-All}$ will need to shrink and will be updated as illustrated in Figures \ref{fig:bounds-check-new}d-\ref{fig:bounds-check-new}e.

Notice that deciding membership of a point into the group requires a constant number of comparisons regardless of the number of points inside Group $g$. Furthermore, the maintenance of the bounding rectangle of the group takes constant time for every inserted point into $g$. Also, notice that $R_{\epsilon-All}$ stops shrinking if its dimensions reach $\epsilon \times \epsilon$, which is a lower-bound on the size of $R_{\epsilon-All}$.
Figure~\ref{fig:bounds-check-new}e gives the updated $R_{\epsilon-All}$ after Point $a_3$ is inserted into the group.

\begin{algorithm}[t]\small
\KwIn{$p_i$: data point, $\epsilon$: similarity threshold , $\delta$: distance function, $CLS$: ON-OVERLAP clause, $G$: set of existing groups }
\KwOut{$CandidateGroups$, $OverlapGroups$}
$CandidateGroups\leftarrow NULL$\\
$OverlapGroups\leftarrow NULL$\\
\For{ each group $g_{j}$ in $G$}{ \uIf {PointInRectangleTest($p_i$, $g_j$) is True}{insert $g_j$ into $CandidateGroups$}
\ElseIf {CLS is not JOIN-ANY \textbf{and} OverlapRectangleTest($p_i$, $g_j$) is True}{
\For{ each $p_k$ in $g_{j}$}{\If { (Distance($p_i$, $p_k$, $\delta$)$\leqslant \epsilon$)}{insert $g_j$ into $OverlapGroups$\\ break}} }}
\caption{\small{Bounds-Checking FindCloseGroups}}
\label{alg:findcloseboundchecking}
\end{algorithm}

\begin{sloppypar}
Procedure~\ref{alg:findcloseboundchecking} gives the pseudocode for \textit{Bounds-Checking FindCloseGroups}.
The procedure uses the $\epsilon$-All bounding rectangle to reduce the number of distance computations needed to realize $FindCloseGroups$ using the $L_{\infty}$ distance metric.
Procedure $PointInRectangleTest$ (Line 4) uses the $\epsilon$-All rectangle to determine in constant time whether $g_j$ is a candidate group for the input point.
Procedure $OverlapRectangleTest$ (Line 6) tests whether the $\epsilon$-All rectangle of $p_i$ overlaps Group $g_j$'s bounding rectangle. In case of an overlap, all data points in $g_j$ are inspected to verify whether the overlap is nonempty.
The correctness of the $\epsilon$-All bounding rectangle for the $L_{\infty}$ distance metric follows from the fact that the rectangles are closed under intersection, i.e., the intersection of two rectangles is also a rectangle.
\end{sloppypar}

\begin{sloppypar}
A major bottleneck of the bounding rectangles approach
is in the need to linearly scan all existing bounding rectangles that represent the groups
to identify sets $CandidateGroups$ and $OverlapGroups$, which is costly.
 To speedup Procedure \textit{Bounds-Checking FindCloseGroups}, we use a spatial access method (e.g., an R-tree ~\cite{BIBExample:guttman1984r}), to index the $R_{\epsilon-All}$ bounding rectangles of the existing groups.

Procedure~\ref{alg:findclose-boundindex} gives the pseudocode for \textit{Index Bounds-Checking FindCloseGroups}. The procedure performs a window query on the index $Groups\_IX$ (Line 4) to retrieve the set $GSet$ of all groups that intersect the bounding rectangle $R_{p_i}$ for the newly inserted point $p_i$. Next, it iterates over $GSet$ (Lines 4-11) and executes $PointInRectangleTest$ to determine whether the inspected group belongs to either one of the two sets $CandidateGroups$ or $OverlapGroups$.
Finally, the elements of $OverlapGroups$ are inspected to retrieve the subset of elements that satisfy the similarity predicate.

\begin{figure}[t]
\centering
\includegraphics[width=2.5in,height=2.2in]{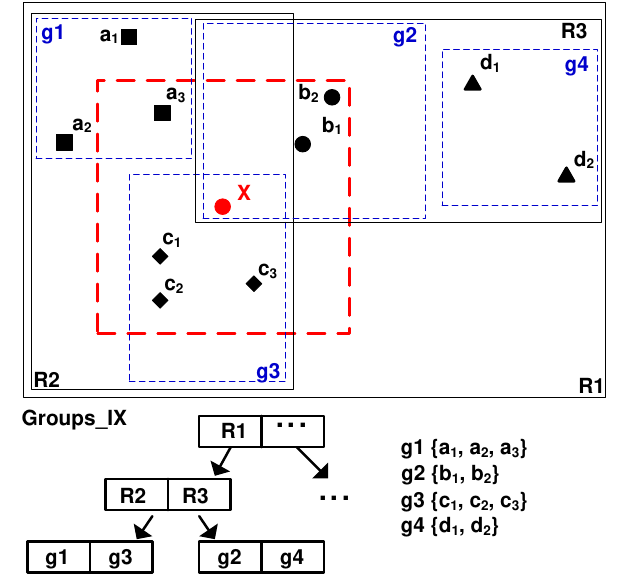}
\caption{SGB-All: performing a window Query on $Groups\_IX$ using $\epsilon = 4$ and $L_{\infty}$}
\label{fig:windowquery}
\end{figure}

Refer to Figure~\ref{fig:windowquery} for illustration.
An R-tree index, termed $Groups\_IX$, is used to index the bounding rectangles of the groups discovered so far.
In this case, $Groups\_IX$ contains bounding rectangles for Groups $g_1$-$g_4$. Given the newly arriving Point $x$, a window query of the $\epsilon$-All rectangle for $x$ is performed on
$Groups\_IX$ that returns all the intersecting rectangles; in this case, $g_1$, $g_2$, and $g_3$. The outcome of the query is used to construct the sets $CandidateGroups$ and $OverlapGroups$.
\end{sloppypar}

\begin{algorithm}[t]\small
\KwIn{$p_i$: data point, $\epsilon$: similarity threshold , $\delta$: distance function, $CLS$: ON-OVERLAP clause, $G$: set of existing groups, $Groups\_IX$: index on $G$'s bounding rectangles }
\KwOut{$CandidateGroups$, $OverlapGroups$}
$CandidateGroups\leftarrow NULL$\\
$OverlapGroups\leftarrow NULL$\\
$R_{p_i} \leftarrow$   CreateBoundingRectangle($p_i$, $\epsilon$)\\
$GSet\leftarrow WindowQuery(p_i, R_{p_i}, Groups\_IX)$\\
\For{ each group $g_{j}$ in $GSet$}{ \uIf {PointInRectangleTest($p_i$, $g_j$) is True}{insert $g_j$ into $CandidateGroups$}
\ElseIf {CLS is not JOIN-ANY}{\For{ each $p_{k}$ in $g_j$}
{\If { (Distance($p_i$, $p_k$, $\delta$)$\leqslant \epsilon$)} {insert $g_j$ into $OverlapGroups$\\ break}}}}
\caption{\small{Index Bounds-Checking FindCloseGroups}}
\label{alg:findclose-boundindex}
\end{algorithm}

\subsection{Handling False Positives $L_2$}
\label{section:false-positive}

\begin{figure}[t]
\centering
\includegraphics[width=2.6in,height=2.3in]{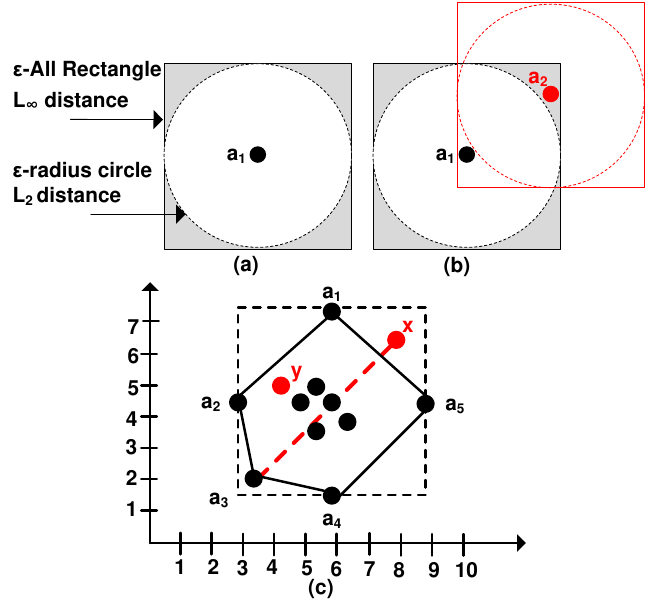}
\caption{(a) The $\epsilon$-radius circle in $L_2$, (b) The problem of false positive for $L_2$, (c) The $\epsilon$-convex hull test for $\epsilon = 6$. }
\label{fig:falsepositive}
\end{figure}

In this section, we study the effect of using $L_2$ as a similarity distance function on the SGB-All operator. Refer to Figure \ref{fig:falsepositive}a for illustration. In contrast to the $L_{\infty}$ distance, the set of points that are exactly $\epsilon$ away from $a_1$ in the $L_2$ metric space form a circle. Inserting $a_2$  (Figure \ref{fig:falsepositive}b) is correct using the $L_{\infty}$ distance since $a_2$ is inside the $\epsilon$-All rectangle of $a_1$'s group. However, under the $L_2$ distance, $a_2$ is more than $\epsilon$ away from $a_1$ since $a_2$ lies outside $a_1$'s $\epsilon$-circle. As a result, all points that are inside $a_1$'s $\epsilon$-All group rectangle but are outside the $\epsilon$-circle (i.e., the grey-shaded area in Figure \ref{fig:falsepositive}b) falsely pass the bounding rectangle test.

 Procedure \textit{Naive FindCloseGroups} in (Procedure \ref{alg:findcloseall-nested}) inspects all input data points. Therefore, the problem of false-positive points does not occur. On the other hand, the Bounds-Checking approach introduced in Procedures \ref{alg:findcloseboundchecking} and \ref{alg:findclose-boundindex} uses the $\epsilon$-All rectangle technique to identify the sets \textit{CandidateGroups} and \textit{OverlapGroups} and hence must address the issue of false-positive points for the $L_2$ distance metric.

We introduce a \textbf{Convex Hull Test} to refine the data points that pass the Bounds-Checking filter step. Given a group of points, a convex hull \cite{BIBExample:de2008computational} is the smallest convex set of points within a group. In Figure \ref{fig:falsepositive}c, the points $a_1$-$a_5$ form the convex hull set for Group $g$. Based on the SGB-All semantics, the diameter of the conevex hull (i.e., the two farthest points) satisfies the similarity predicate.

\begin{algorithm}[t]\small
\KwIn{$p_i$: data point, $g$: existing group }
\KwOut{True if $p_i$ is not false positive, False otherwise}
\SetAlgoNoLine
 $ConvexHullSet\leftarrow$  $getConvexHull(g)$\\
 \uIf {$p_i$ inside convex hull}{return True}
 \Else { $farthestPoint\leftarrow$  $getMaxDistElem(ConvexHullSet,p_{i})$\\
 \If{$distance(farthestDistPoint,p_{i} ) <= \epsilon$}{ return True}}
  return False
\caption{\small{Convex Hull Test}}
\label{alg:convexhulltest}
\end{algorithm}

The \textit{Convex Hull Test}, illustrated in Procedure \ref{alg:convexhulltest}, verifies whether a point is a false-positive. This additional test can be inserted immediately after (Line 4) in
Procedure~\ref{alg:findcloseboundchecking} or immediately after (Line 6) in
Procedure~\ref{alg:findclose-boundindex}. Consequently, any new point that lies inside a group's convex hull (e.g., Point $y$ in Figure~\ref{fig:falsepositive}c) satisfies the similarity predicate. In addition, in order to verify points that are outside the convex hull (e.g., Point $x$ in Figure \ref{fig:falsepositive}c), it is enough to evaluate the similarity predicate between $p_i$ and the convex hull. The correctness of the convex hull test follows from the fact that the convex hull set contains the farthest point from $p_i$, say $p_{f}$. Therefore, it is sufficient to evaluate the similarity predicate between $p_i$ and $p_f$ (e.g., Point $x$ and Point $a_3$ in Figure \ref{fig:falsepositive}c). Section \ref{section:complexity analysis} discusses the complexity of the convex hull approach.

\section{Algorithms for SGB-Any}
\label{section:sgb-any-framework}
\begin{sloppypar}

In this section, we present an algorithmic framework to realize similarity-based grouping using the distance-to-any semantics. The generic SGB-Any framework in Procedure~\ref{alg:SGBAny} proceeds as follows. For each data point, say $p_i$, Procedure $FindCandidateGroups$ (Line 2) uses the distance-to-any similarity predicate to identify the set $CandidateGroups$ that consists of all the existing groups that $p_i$ can join. In contrast to SGB-All, in the
distance-to-any semantics, a point, say $p_i$, can join a candidate group, say $g$, when $p_i$ is within a predefined similarity threshold from at least one another point in $g$. Procedure $ProcessGroupingANY$ (Line 3) inserts $p_i$ into a new or an existing group.
\end{sloppypar}

\begin{algorithm}[t]\small
\KwIn{$P$: set of data points, $\epsilon$: similarity threshold, $\delta$: distance function, $Points\_IX$: spatial index}
\KwOut{Set of groups G}
\For{ each data element $p_{i}$ in $P$}
{
 $CandidateGroups \leftarrow FindCandidateGroups(p_i, $Points\_IX$, \epsilon, \delta)$\\
 $ProcessGroupingANY(p_i, CandidateGroups)$\\

}
\caption{\small{Similarity Group-By ANY Framework}}
\label{alg:SGBAny}
\end{algorithm}


\subsection{Finding Candidate Groups}
\begin{sloppypar}
A Naive \textit{FindCandidateGroups} approach similar to Procedure 2 can identify the set $CandidateGroups$. However, this solution incurs many distance computations, and brings the upper-bound time complexity of the SGB-Any framework to $O(n^2)$. The filter-refine paradigm using an $\epsilon$-group bounds-checking approach while applying a distance-to-any predicate (i.e., similar to Procedures \ref{alg:findcloseboundchecking}-\ref{alg:convexhulltest}) suffers from two main challenges.
By drawing squares of size $\epsilon\times \epsilon$ around the input point and forming a bounding rectangle that encloses all these squares results in a consecutive chain-like region and the area of false-positive
progressively increases in size as we add new data points.
Furthermore, the convex hull approach to test for false-positive points cannot be applied in SGB-Any as it suffers from false-negatives caused by the fact that the length of the diameter of the convex hull can actually be more than $\epsilon$ in the case of SGB-Any.  Details are omitted here for brevity.

\begin{figure}[t]
\centering
\includegraphics[width=3.3in,height=2.2in]{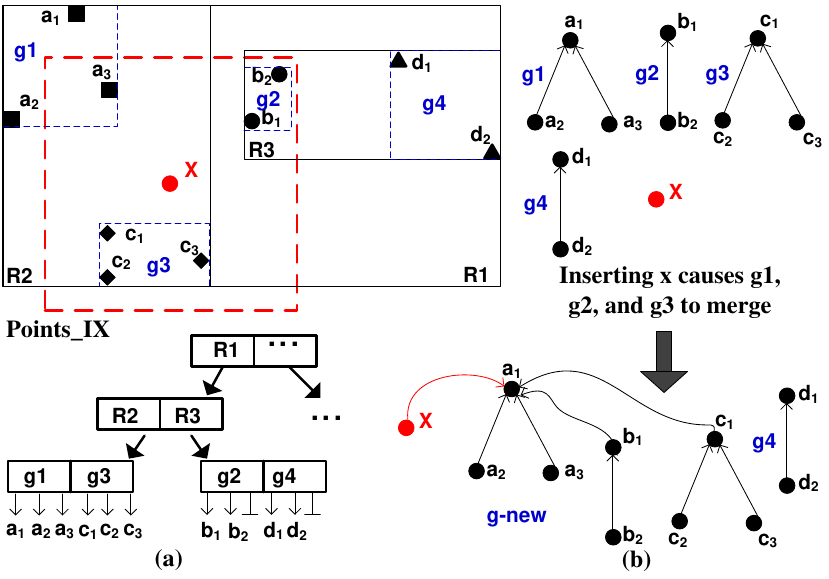}
\caption{(a) SGB-Any: Performing a window query on $Points\_IX \epsilon=4$ using $L_\infty$ (b) The disjoint data structure Union-Find is used to maintain existing groups.}
\label{fig:sgbanyindex-unionfind}
\end{figure}

Consequently, \textit{FindCandidateGroups} in Procedure~\ref{alg:findcloseany} uses an R-tree index, termed $Points\_IX$. $Points\_IX$ maintains the previously processed data points to efficiently find $CandidateGroups$.
Refer to Figure~\ref{fig:sgbanyindex-unionfind} for illustration.
For an incoming point, say Point $x$, 
an $\epsilon$-rectangle (Line 2 of Procedure~\ref{alg:findcloseany}) is created to perform a window query on $Points\_IX$ to retrieve $PointsSet$ (Line 3). $PointsSet$ corresponds to the points that are within $epsilon$ from $x$, e.g., $\{a_3, c_1, c_2, c_3, b_1, b_2\}$. Based on the semantics of SGB-Any, $CandidateGroups$ contains the groups that cover the points in $PointsSet$. For instance, point $a_3$ belongs to $g_1$, points $\{c_1  \textendash c_3\}$ belong to $g_2$, and points $\{b_1 \textendash b_2\}$ belong to group $g_3$. Hence, $CandidateGroups = \{g_1, g_2, g_3\}$.
Procedure $GetGroups$ (Line 7) employs a Union-Find data structure~\cite{BIBExample:Tarjan} to keep track of existing, newly created, and merged groups (see Figure~\ref{fig:sgbanyindex-unionfind}b) to efficiently construct $CandidateGroups$ given $PointsSet$.

\end{sloppypar}

\begin{algorithm}[t]\small
\KwIn{$p_i$: data point, $Points\_IX$: spatial index, $\delta$: distance function, $\epsilon$: similarity threshold}
\KwOut{$CandidateGroups$}
$CandidateGroups\leftarrow NULL$\\
$R_{pi} \leftarrow$   CreateBoundingRectangle($p_i$, $\epsilon$)\\
$PointsSet\leftarrow WindowQuery(p_i, R_{pi}, Points\_IX)$\\
\If{$\delta$ is $L_2$}{$PointsSet\leftarrow VerifyPoints(Points\_IX, \delta, \epsilon)$}
$CandidateGroups\leftarrow GetGroups(PointsSet)$\\
insert $p_i$ into $Points\_IX$
\caption{\small{FindCandidateGroups}}
\label{alg:findcloseany}
\end{algorithm}

\subsection{Processing New Points}
\begin{sloppypar}

Procedure~\ref{alg:ProcessGroupingANY} gives the pseudocode for \textit{ProcessGroupingANY}.
Lines~1-6 identify the cases when $CandidateGroups$ is empty, or when it consists of one group. In these cases, $p_i$ is inserted into a newly created group or into an existing group. Next, it handles the case that occurs when $p_i$ is close to more than one group. In the SGB-Any semantics, all candidate groups that $p_i$ can join are merged into one group. Therefore, Procedure $MergeGroupsInsert$ (Line 8) handles merging candidate groups and then inserts $p_i$ into the merged groups.
Referring to Figure~\ref{fig:sgbanyindex-unionfind}b, Point $x$ overlaps groups $g_1, g_2,$ and $g_3$. Based on the semantics of SGB-Any, the overlapped groups $g_1$, $g_2$, and $g_3$ are merged into one encompassing bigger group, termed $G \textendash new$. In this case, the root pointers of $g_1$, $g_2$ and $x$ in the Union-Find data strucure are redirected to Point $a_1$.

\end{sloppypar}

\begin{algorithm}[t]\small
\KwIn{$p_i$: data point, $CandidateGroups$ }
\KwOut{updates $CandidateGroups$}
\uIf{$CandidateGroups$ is Empty}
 {create a new group $g_{new}$ \\
 $ProcessInsert(p_i, g_{new})$}
 \uElseIf {sizeof(CandidateGroups) == 1 }{ insert into existing group $g_{out}$\\  $ProcessInsert(p_i, g_{out})$ }
 \Else {MergeGroupsInsert($CandidateGroups$, $p_i$)}
\caption{\small{ProcessGroupingANY} }
\label{alg:ProcessGroupingANY}
\end{algorithm}

\section{Complexity, Realization, and \\
Evaluation}
\label{section:performance-evaluation}

\subsection{Complexity Analysis}
\label{section:complexity analysis}

\begin{sloppypar}
Table \ref{SGBAll_table_complexity} summarizes the average-case running time of SGB-All using the proposed optimizations for the $L_{\infty}$ distance metric. The \textit{All-Pairs} algorithm corresponds to naive \textit{FindCloseGroups} in Procedure \ref{alg:sgball-framework}. Similarly, \textit{Bounds-Checking} and \textit{On-the-fly Indexing} corresponds to the \textit{Bounds-Checking} and \textit{Index Bounds-Checking} optimizations, where $|G|$ is the number of output Groups and $m$ is the recursion depth for the ON-OVERLAP FORM-NEW. In addition, the average-case running time of SGB-Any when using the index is $O(n\log n)$.  The worst-case and best-case running times, and detailed analysis are given in the Appendix.
\end{sloppypar}

\begin{sloppypar}

\begin{table}[h]
    \centering
    \scriptsize\addtolength{\tabcolsep}{-5pt}
        \begin{tabular}{|c|c|c|c|}
        \hline
                & JOIN-ANY & ELIMINATE & FORM-NEW-GROUP \\\hline
                All-Pairs & $O(n^2)$& $O(n^2)$& $O(n^3)$ \\\hline
                Bounds-Checking & $O(n|G|)$& $O(n|G|))$& $O(mn|G|) $ \\\hline
                on-the-fly Index & $O(n\log|G|)$& $O(n\log|G|)$& $O(mn\log|G|)$ \\\hline
        \end{tabular}
        \caption{SGB-All Complexity for the $L_{\infty}$ distance}
\label{SGBAll_table_complexity}
\end{table}

\end{sloppypar}

\begin{sloppypar}
\begin{table}[h]
\tiny
    \centering
    \scriptsize\addtolength{\tabcolsep}{-6pt}
        \begin{tabular}{|c|l|}
        \hline
        \multicolumn{2}{|l|}{\textbf{Business Question}: Retrieve large volume customers} \\
        \hline
         GB1 & Same as the TPCH-Q18 \\
          \hline
        \multicolumn{2}{|l|}{\tabincell{l} {\textbf{Business Question}: Retrieve customers with similar buying power, account balance}} \\
        \hline
                 \tabincell{l}{SGB1 \\or \\ SGB2} & \tabincell{l} {
                 SELECT max(ab), min(tb),max(tb), average(ab), array\_agg(R1.c\_custkey) \\
                  \quad \ \quad FROM (SELECT c\_custkey, c\_acctbal as ab FROM Customer  \\
                  \quad \ \quad WHERE c\_acctbal $>$100 ) as R1 \\
                 \quad \ \quad (SELECT o\_custkey,  sum(o\_totalprice) as tp FROM  Orders, Lineitem    \\
                 \quad \ \quad   WHERE o\_orderkey  in (SELECT l\_orderkey FROM  lineitem  \\
                  \quad \ \quad  GROUP BY Rl\_orderkey having sum(l\_quantity) $>$3000)  \\
                \quad \ \quad  and o\_orderkey =l\_orderkey and  o\_totalprice $>$ 30000) as R2  \\
                \quad \ \quad WHERE R1.c\_custkey=R2.o\_custkey \\
                 GROUP BY ab,tp \textbf{DISTANCE-ALL} WITHIN $\epsilon$  USING lone/ltwo \\
                 on\_overlap join-any/form-new/eliminate \\
                 \textbf{or} GROUP BY ab,tp \textbf{DISTANCE-ANY} WITHIN $\epsilon$ USING lone/ltwo
               } \\
       \hline
       \hline
        \multicolumn{2}{|l|}{\tabincell{l} {\textbf{Business Question}: \\
        Report profit on a given line of parts (by supplier nation and year)}}\\
        \hline
          GB2 & Same as the TPCH-Q9\\
          \hline
        \multicolumn{2}{|l|}{\tabincell{l}{\textbf{Business Question}:
        \\Report profit and shipment time of parts share similar profit and shipment date}}  \\
        \hline
          \tabincell{l}{SGB3 \\or \\ SGB4} &  \tabincell{l} {
          SELECT count(\*),sum(tprof), sum(stime) FROM \\
           \quad \ \quad (SELECT  ps\_partkey as partkey, sum(l\_extendedprice * (1 - l\_discount) \\
           \quad \ \quad  - ps\_supplycost *l\_quantity)  as tprof, sum(l\_receiptdate-l\_shipdate) \\
           \quad \ \quad as stime FROM  lineitem, partsupp,supplier WHERE  ps\_partkey = \\
            \quad \ \quad   l\_partkey and  s\_suppkey=ps\_suppkey GROUP BY ps\_partkey) as profit \\
             GROUP BY tprof, stime \textbf{DISTANCE-ALL} WITHIN $\epsilon$  USING lone/ltwo \\
             on\_overlap join-any/form-new/eliminate \\
             \textbf{or} GROUP BY tprof, stime \textbf{DISTANCE-ANY} WITHIN $\epsilon$ USING lone/ltwo
          } \\
        \hline
        \hline
         \multicolumn{2}{|l|}{\tabincell{l} {\textbf{Business Question}: \\
        Determines top supplier who contributed the most to the overall revenue for parts)}}\\
        \hline
          GB3 & Same as the TPCH-Q15\\
          \hline
        \multicolumn{2}{|l|}{\tabincell{l}{\textbf{Business Question}:
        \\Report supplier who contributed the similar profit and account balance}}  \\
        \hline
        \tabincell{l}{SGB5 \\or \\ SGB6} &  \tabincell{l} {
          SELECT array\_agg(s\_suppkey), sum(r.trevenue), sum(s\_acctbal)  \\
           \quad \ \quad   FROM (SELECT l\_suppkey as suppkey, sum(l\_extendedprice * (1 -  \\
           \quad \ \quad    l\_discount)) as trevenue , sum(s\_acctbal) As acctbal  FROM  Lineitem \\
           \quad \ \quad  WHERE l\_shipdate  $>$ date '[1995-01-01]' and l\_shipdate $<$ date \\
           \quad \ \quad '[1996-01-01]'+ interval '10' month GROUP  BY l\_suppkey )as r \\
             GROUP BY r.trevenue, s\_acctbal  \textbf{DISTANCE-ALL} WITHIN $\epsilon$ \\
              USING lone/ltwo on\_overlap join-any/form-new/eliminate \textbf{or} GROUP \\
              BY r.trevenue, s\_acctbal \textbf{DISTANCE-ANY} WITHIN $\epsilon$ USING lone/ltwo
          } \\
        \hline
        \end{tabular}
        \caption{Performance Evaluation Queries on TPC-H}
\label{tab-queries}
\end{table}
\end{sloppypar}

\subsection{Implementation}

\begin{sloppypar}
We realize the proposed SGB operators inside PostgreSQL.
In the \textit{parser}, the grammar rules, and actions related to the ``SELECT" statement syntax are updated with similarity keywords (e.g., DISTANCE-TO-ALL and DISTANCE-TO-ANY) to support the SGB query syntax. The parse and query trees are augmented with parameters that contain the similarity semantics (e.g., the threshold value and the overlap action). The \textit{Planner and Optimizer} routines use the extended query-tree to create a similarity-aware plan-tree.
In this extension, the optimizer is manipulated to choose a hash-based SGB plan.

The executor modifies the hash-based aggregate group-by routine.
Typically, an aggregate operation is carried out by the incremental evaluation of the aggregate function on the processed data. However, the semantics of ON-OVERLAP ELIMINATE and ON-OVERLAP FORM-NEW-GROUP can realize final groupings only after processing the complete dataset. Therefore, the aggregate hash table keeps track of the existing groups in the following way. First, the aggregate hash table entry (AggHashEntry) is extended with a TupleStore data structure that serves as a temporary storage for the previously processed data points. Next, referring to the Bounds-Checking FindCloseGroups presented in Procedure \ref{alg:findcloseboundchecking}, each group's bounding rectangle is mapped into an entry inside the hash directory. Bounds-Checking FindCloseGroups linearly iterates over the hash table directory to build the sets $CandidateGroups$ and $OverlapGroups$. The Index Bounds-Checking in Procedure \ref{alg:findclose-boundindex} employs a spatial index to efficiently look up all existing groups a data point can join. Consequently, we extend the executor with an in-memory R-tree that efficiently indexes the existing groups' bounding rectangles.

In the implementation of FindCloseGroupsAny in Procedure \ref{alg:findcloseany}, a spatial index is created to maintain the set of points that have been processed and assigned to groups. Moreover, we extend the executor with the Union-Find data structure Disjoint-set forest to support the operations $GetGroups$ and $MergeGroupsInsert$.
\end{sloppypar}


\begin{figure*}
        \centering
        \begin{subfigure}[b]{0.23\textwidth}
                \includegraphics[width=\textwidth]{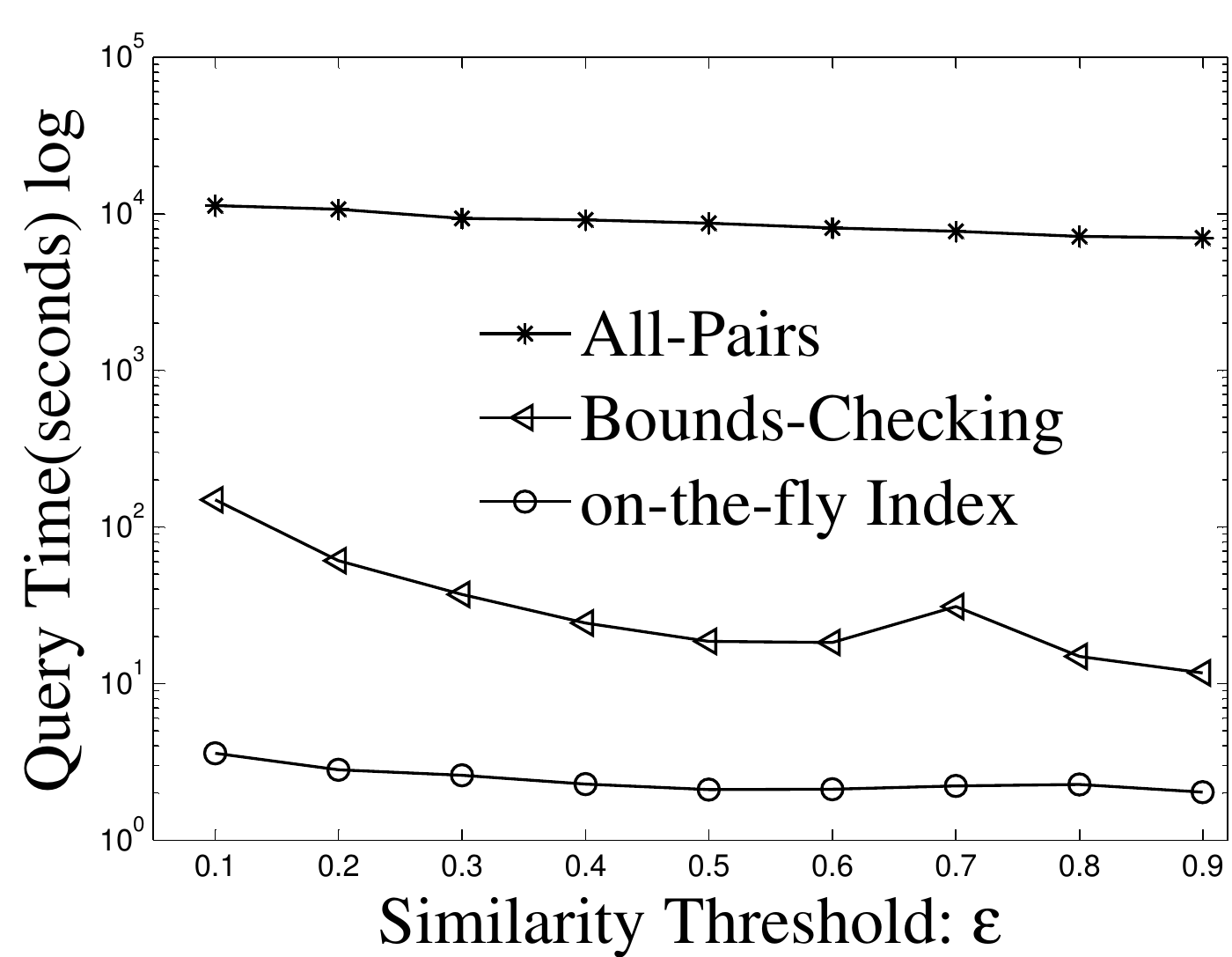}
                \caption{\tiny{SGB-All:JOIN-ANY}}
                \label{fig:SGBALL-JOINANY_EPS}
        \end{subfigure}%
        ~ 
        \begin{subfigure}[b]{0.23\textwidth}
                \includegraphics[width=\textwidth]{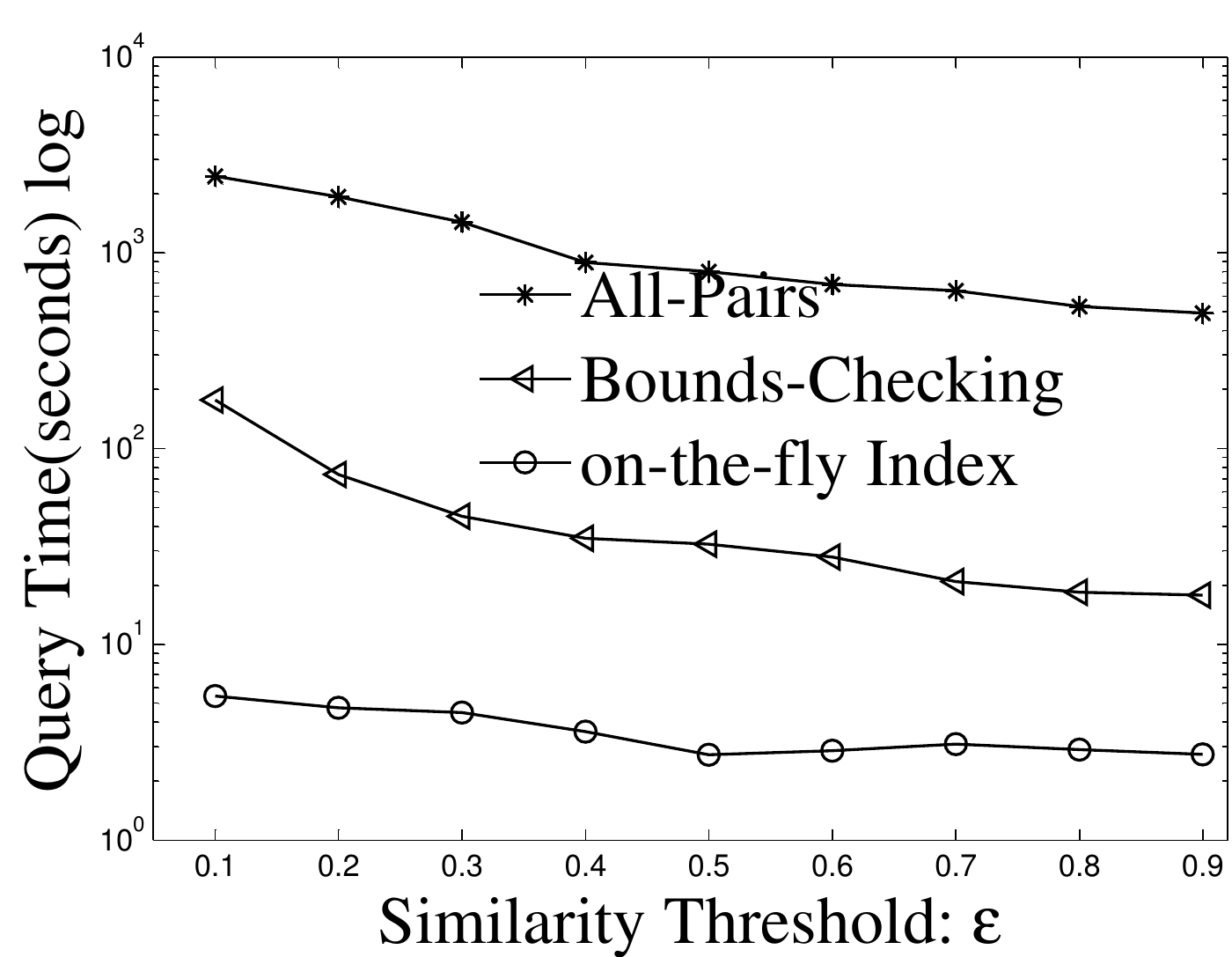}
                \caption{\tiny{SGB-All:ELIMINATE}}
                \label{fig:SGBALL-ELIMINATE_EPS}
        \end{subfigure}
        \begin{subfigure}[b]{0.23\textwidth}
                \includegraphics[width=\textwidth]{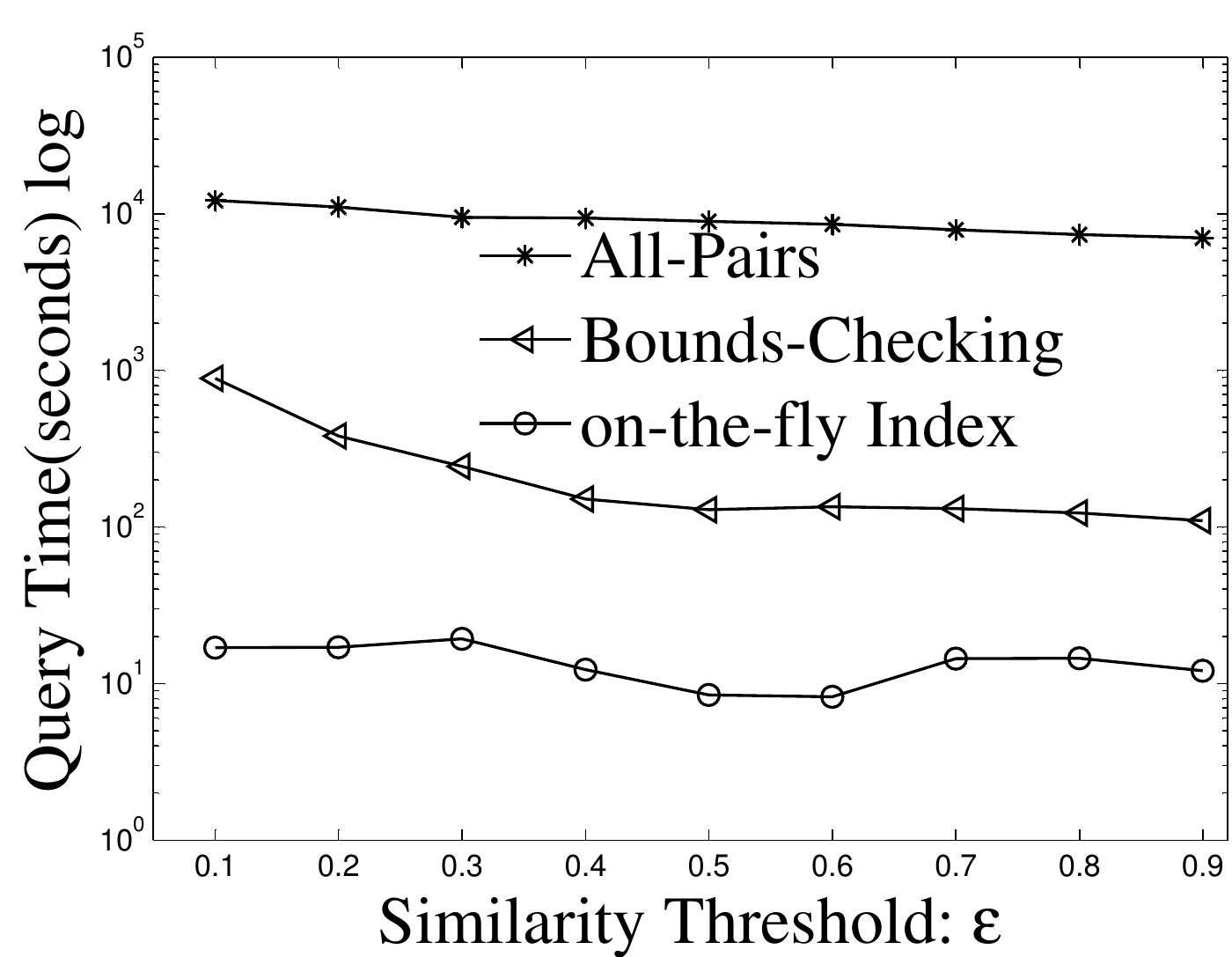}
                \caption{\tiny{SGB-All:FORM-NEW-GROUP} }
                \label{fig:SGBALL-FORMNEW_EPS}
        \end{subfigure}
         \begin{subfigure}[b]{0.23\textwidth}
                \includegraphics[width=\textwidth]{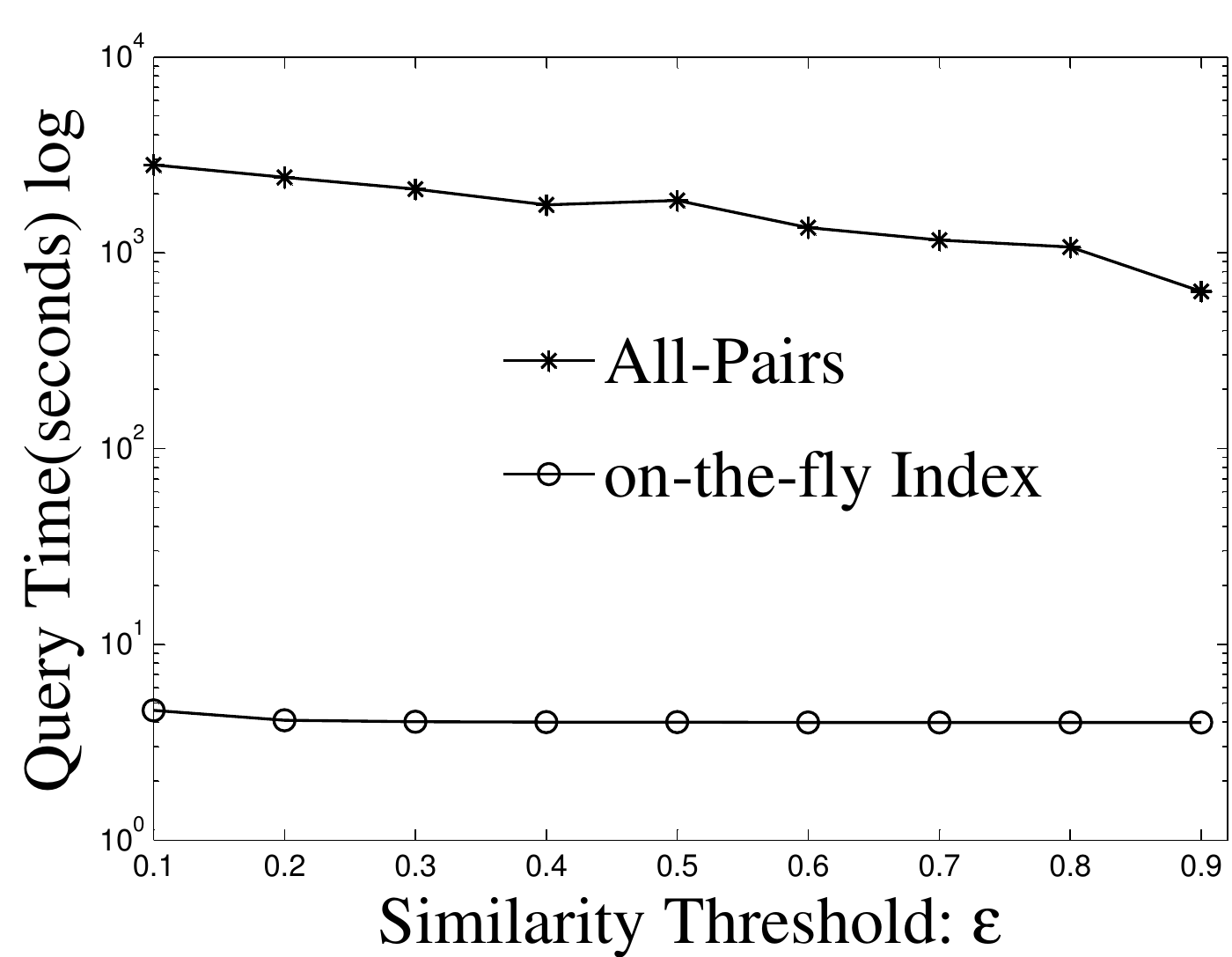}
                \caption{\tiny{SGB-ANY}}
                \label{fig:SGB-ANY_EPS}
        \end{subfigure}
        \caption{The effect of similarity threshold $\epsilon$ on the SGB-All variants and SGB-ANY}\label{fig:effect-eps}
\end{figure*}

\begin{figure*}
        \centering
        \begin{subfigure}[b]{0.23\textwidth}
                \includegraphics[width=\textwidth]{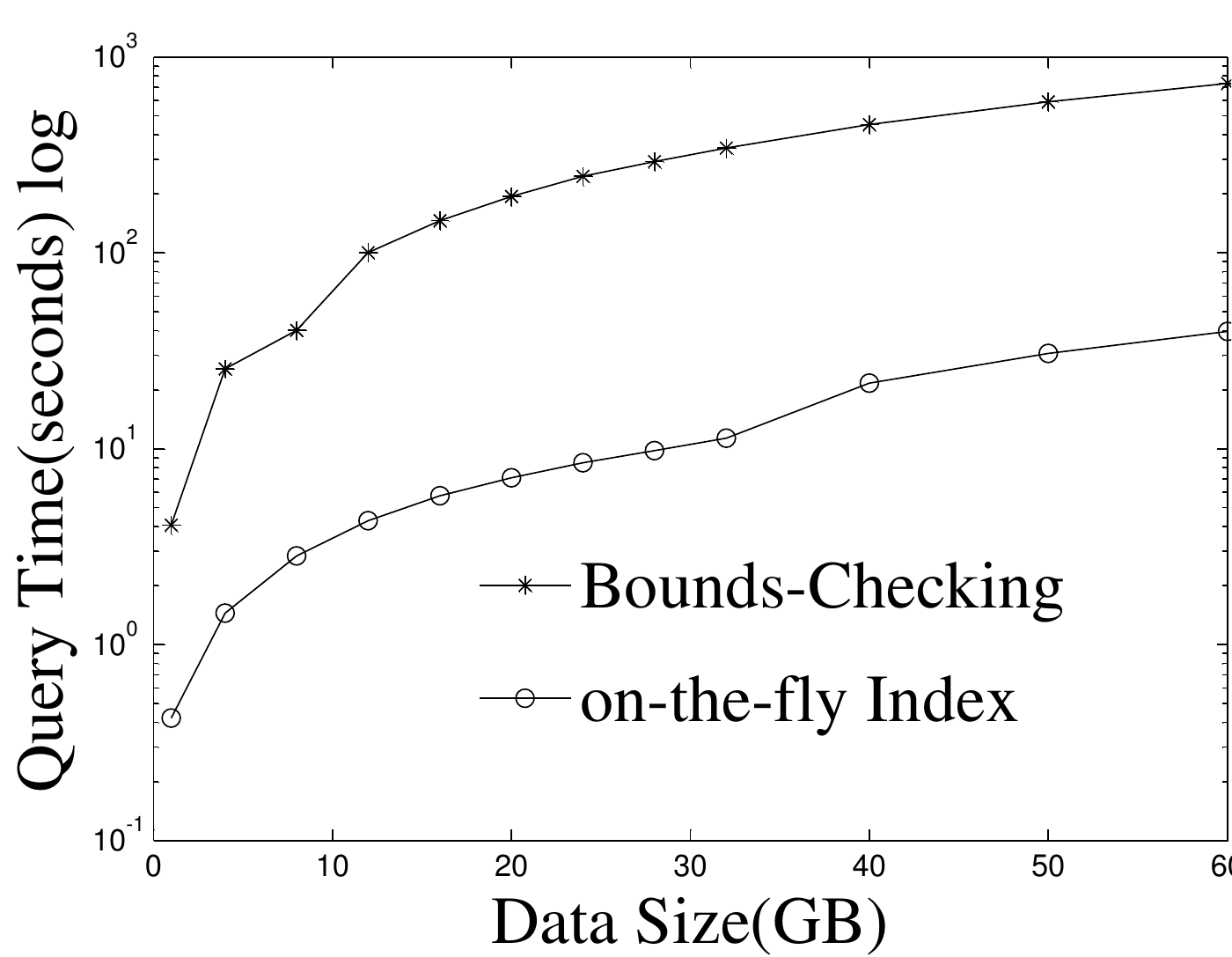}
                \caption{\tiny{SGB-All:JOIN-ANY}}
                \label{fig:SGBALL-JOINANY_DS}
        \end{subfigure}%
        ~ 
        \begin{subfigure}[b]{0.23\textwidth}
                \includegraphics[width=\textwidth]{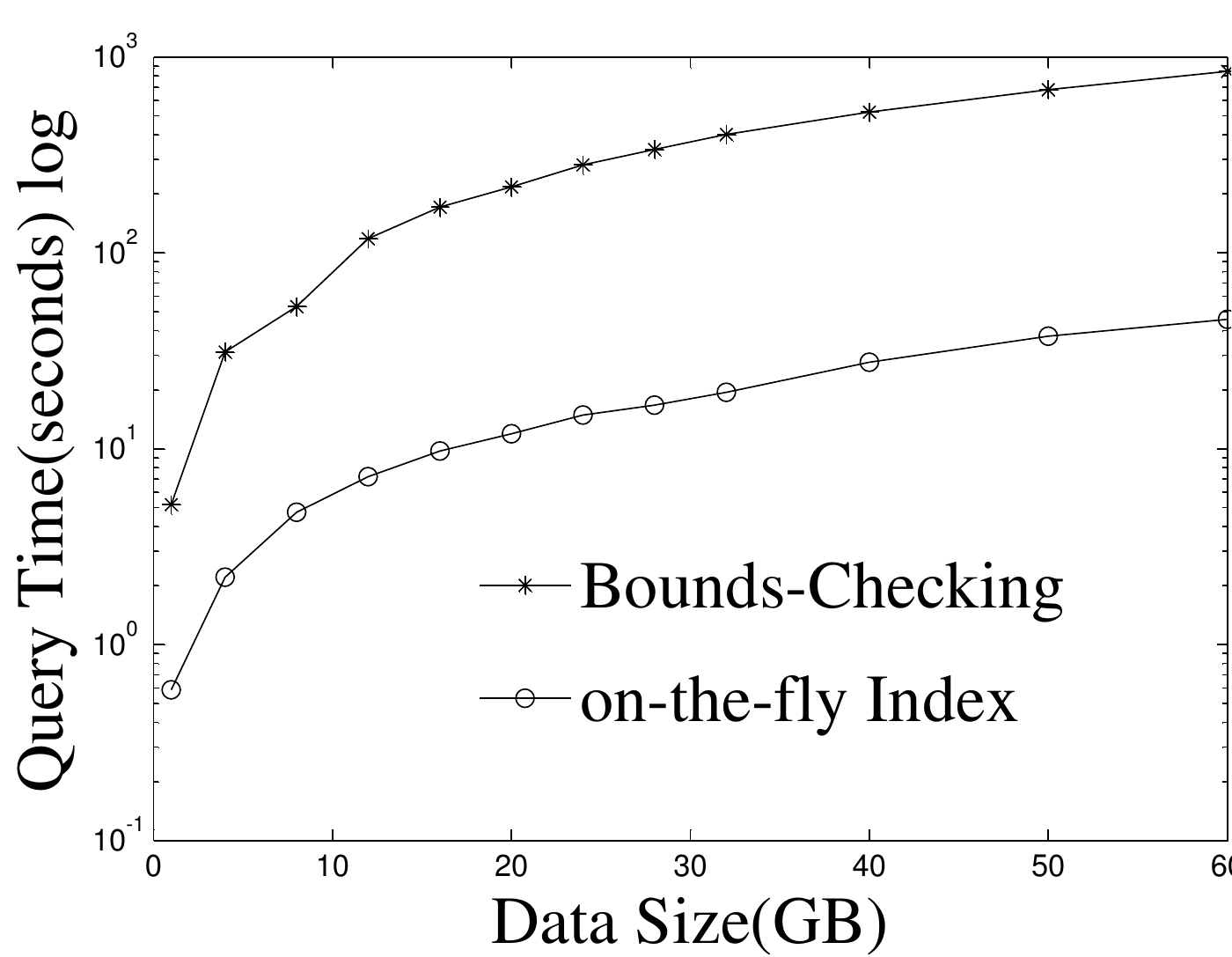}
                \caption{\tiny{SGB-All:ELIMINATE}}
                \label{fig:SGBALL-ELIMINATE_DS}
        \end{subfigure}
        \begin{subfigure}[b]{0.23\textwidth}
                \includegraphics[width=\textwidth]{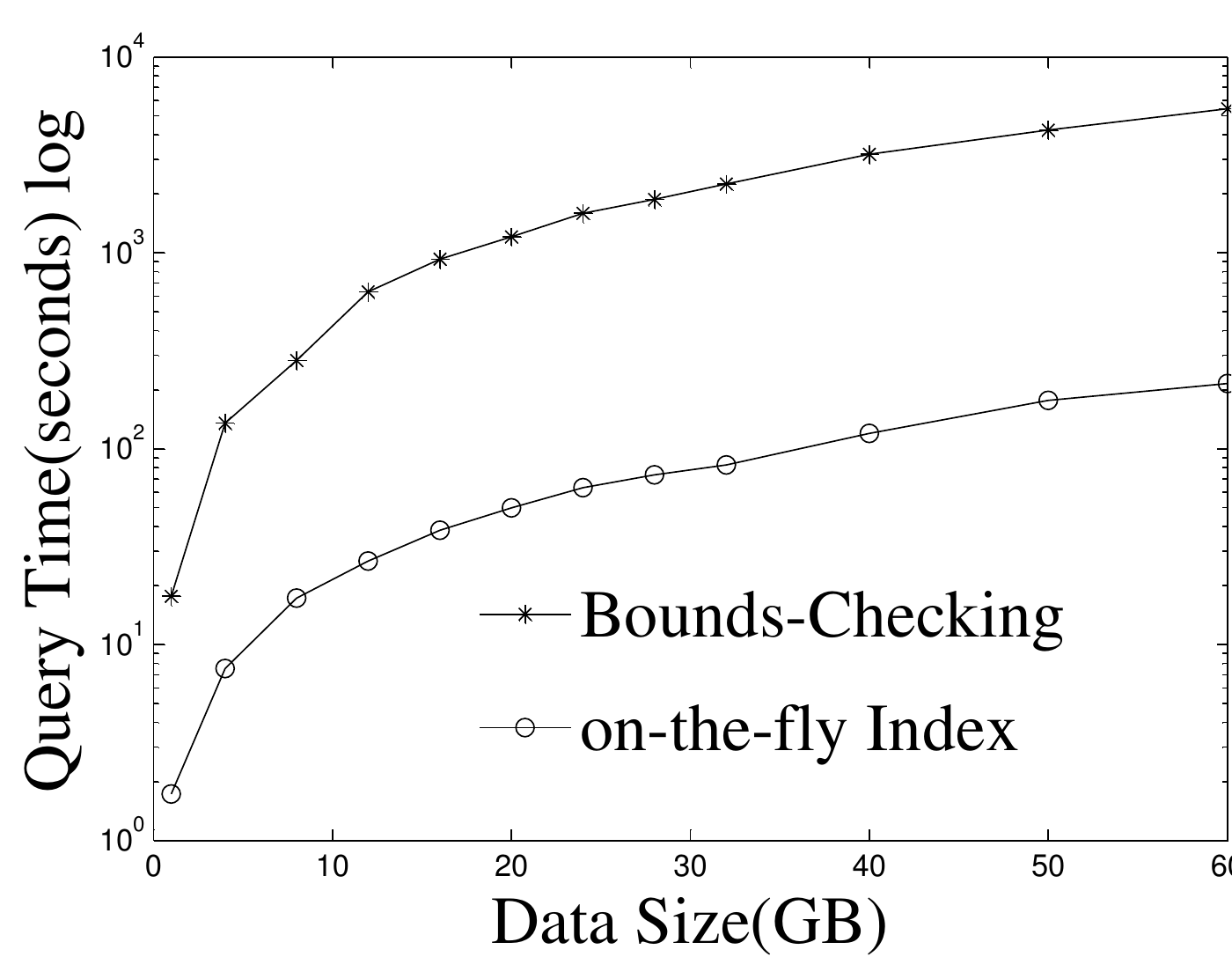}
                \caption{\tiny{SGB-All:FORM-NEW-GROUP} }
                \label{fig:SGBALL-newgroup_DS}
        \end{subfigure}
         \begin{subfigure}[b]{0.23\textwidth}
                \includegraphics[width=\textwidth]{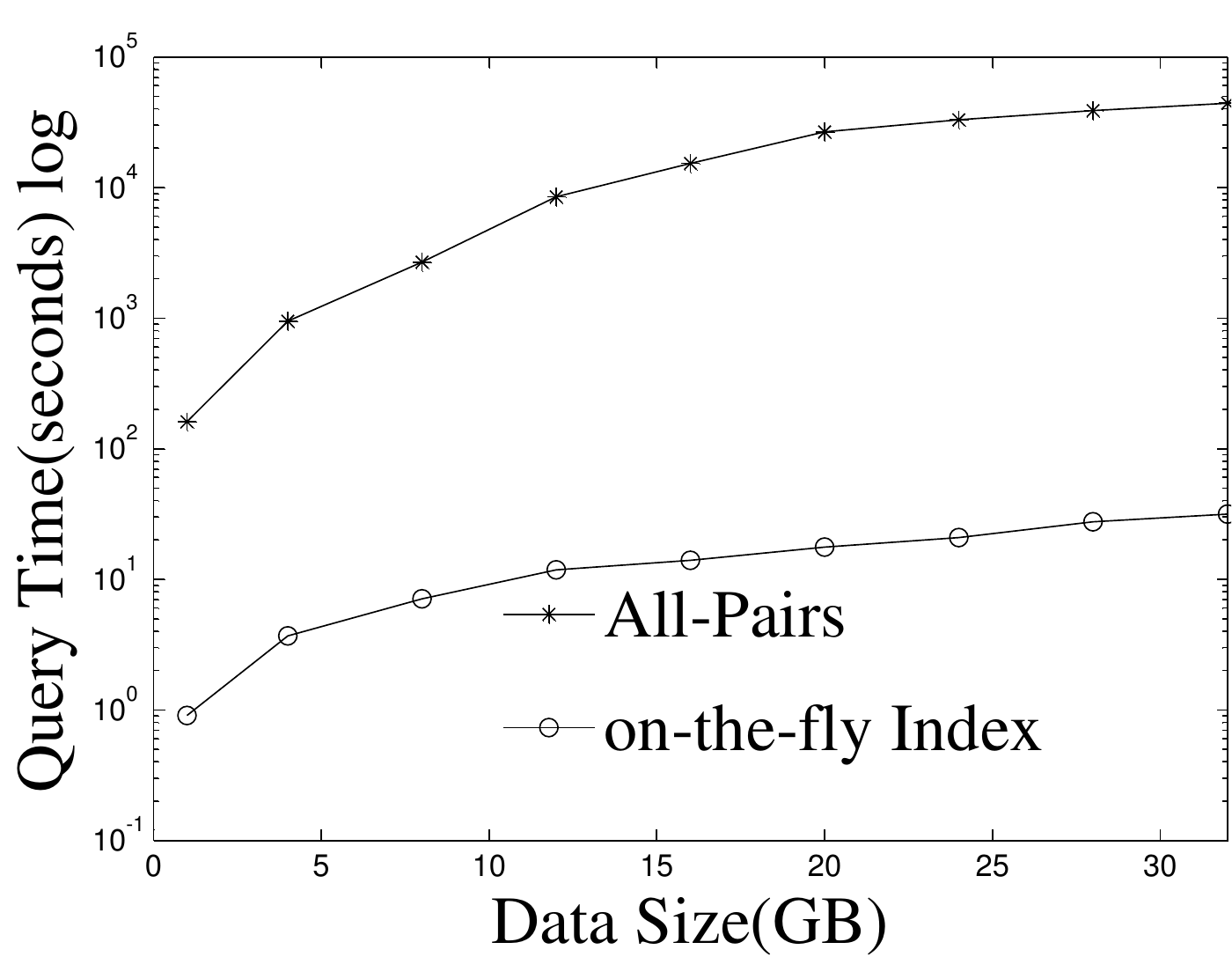}
                \caption{\tiny{SGB-ANY}}
                \label{fig:SGB_ANY_DS}
        \end{subfigure}
        \caption{The effect of increasing data size on the SGB-All variants and SGB-ANY}\label{fig:effect-datasize}
\end{figure*}

\subsection{Datasets}
\begin{sloppypar}

The goal of the experimental study is to
validate the effectiveness of the proposed SGB-All and SGB-Any operators using the optimization
methods discussed in Sections~\ref{section:sgball-framework} and~\ref{section:sgb-any-framework}.
The datasets used in the experiments are based on the TPC-H benchmark\footnote{http://www.tpc.org/tpch/}~\cite{BIBExample:softonline},
and two real-world social checking datasets, namely Brightite\footnote{https://snap.stanford.edu/data/loc-brightkite.html} and Gowalla\footnote{https://snap.stanford.edu/data/loc-gowalla.html} ~\cite{standford_socialdata_2011}. Table~\ref{tab-queries} shows the queries used for performance evaluation experiments on TPC-H data. The multi-dimensional attribute is the combination of different tables. For example, SGB queries, i.e., SGB1/SGB2, are combination of Customer and Order Table, and the number of tuples in the Customer and Order tables is $150K\times SF$ and $1500K\times SF$, respectively, where the scale factor $SF$ ranges from 1 to 60.
For Brightite and Gowalla data, SGB queries follow
Queries~\ref{query:manet-any} and~\ref{query:social-sgball} to cluster users into groups by the corresponding users' check-in information (i.e., latitude and longitude).

The experiments are performed on an Intel(R) Xeon (R) E5320 1.86 GHz 4-core processor with 8G memory running Linux,
and using the default configuration parameters in PostgreSQL. At first, we focus on the time taken by SGB and hence disregard the data preprocessing time, (e.g., the inner join and filter predicates in Query 18).
Furthermore, to understand the overhead of new SGB query, we calculate SGB response time with complicated queries (e.g., the SGB Query 3 to 6). In the paper,
we only give the execution time of the $L_{2}$ distance metric because the performance when using the $L_\infty$ distance metric exhibits a similar behavior.
\end{sloppypar}

\subsection{Effect of similarity threshold $\epsilon$}
\begin{sloppypar}

The effect of the similarity threshold $\epsilon$ on the query runtime is given in Figure~\ref{fig:effect-eps} for SGB-Any and all three overlap variants of SGB-All; JOIN-ANY, ELIMINATE and FORM-NEW-GROUP. The experimental data consists of 0.5 million records. The similarity threshold $\epsilon$ varies from 0.1 to 0.9.

Consider an unskewed dataset, performing SGB-All using a smaller value of $\epsilon$ (e.g., 0.1 or 0.2) forms too many output groups because the similarity predicate evaluates to true on small groups of the data.
Increasing the value of $\epsilon$ forms large groups that decreases the expected number of output groups. Thus, we observe in Figure~\ref{fig:SGBALL-JOINANY_EPS},~\ref{fig:SGBALL-ELIMINATE_EPS},~\ref{fig:SGBALL-FORMNEW_EPS} that the runtime of SGB-All using the various semantics decreases as the value of $\epsilon$ approaches 0.9 with the exception of $\epsilon$ of value 0.7. The slight increase in runtime in the JOIN-ANY and FORM-NEW-GROUP semantics can be attributed to the distribution of the experimental data.

The runtime and speedup in Figure~\ref{fig:SGBALL-JOINANY_EPS},~\ref{fig:SGBALL-ELIMINATE_EPS},~\ref{fig:SGBALL-FORMNEW_EPS} validate the advantage of the optimizations for \textit{Bounds-Checking} and \textit{on-the-fly Index} over \textit{All-Pairs}. The \textit{on-the-fly Index} approach shows two orders of magnitude speedup over \textit{All-Pairs}, and  \textit{Bounds-Checking} approach wins one order magnitude faster than that of \textit{All-Pairs}. The reason is that \textit{All-Pairs} realizes similarity grouping by inspecting all pairs of data points in the input, and its runtime is bounded by the input size. In contrast, \textit{Bounds-Checking} defines group bounds in conjunction with the similarity threshold to avoid excessive runtime while grouping. Therefore, the runtime of \textit{Bounds-Checking} is bounded by the number of output groups. Lastly, indexing output groups using \textit{on-the-fly Index} alleviates the effect of the number of output groups on the overall runtime and makes it steady across the various ON-OVERLAP options.

The effect of the similarity threshold $\epsilon$ on the query runtime for the SGB-Any query is given in Figure~\ref{fig:SGB-ANY_EPS}.
The experiment illustrates that the runtime for \textit{All-Pairs} SGB-Any decreases as the value of $\epsilon$ increases.
Furthermore, the runtime of the \textit{on-the-fly Index} method slightly changes. As a result, the speedup between the \textit{All Pairs} and the \textit{on-the-fly Index} methods slightly decreases.
The runtime result validates that the performance of the \textit{on-the-fly Index} method is stable as we vary the value of $\epsilon$. The reason is that the Union-Find data structure efficiently finds and merges the candidate groups. Figure~\ref{fig:SGB-ANY_EPS} verifies that, for all values of $\epsilon$, the runtime performance of the \textit{on-the-fly Index} method for SGB-Any is two orders of mangitude faster than the \textit{All-Pairs} SGB-Any.
\end{sloppypar}

\subsection{Speedup}

\begin{sloppypar}
Figure~\ref{fig:SGBALL-JOINANY_DS},~\ref{fig:SGBALL-ELIMINATE_DS} and~\ref{fig:SGBALL-newgroup_DS} give the performance and speedup of the \textit{Bounds-Checking} and \textit{on-the-fly Index}
methods for large datasets with scale factor up to 60. The similarity threshold $\epsilon$ is fixed to 0.2. We do not show the results for the naive approach \textit{All-Pairs} because its runtime increases quadratically as the data size increases. From Figure~\ref{fig:SGBALL-JOINANY_DS},~\ref{fig:SGBALL-ELIMINATE_DS} and~\ref{fig:SGBALL-newgroup_DS}, we observe that the runtime of the \textit{Bounds-Checking} method increases as the number and size of groups increases. The \textit{on-the-fly Index Bounds-Checking} method finds the sets $CandidateGroups$ and $OverlapGroups$ efficiently using the R-tree index, and the runtime of \textit{on-the-fly Index Bounds-Checking} method increases steadily and is consistently lower than the \textit{Bounds-Checking} methods. We observe that the speedup of the \textit{on-the-fly Index Bounds-Checking} method is one order of magnitude better than that of \textit{Bounds-Checking}.

Figure~\ref{fig:SGB_ANY_DS} gives the effect of varying the data size on the runtime of SGB-Any when $\epsilon$ is fixed to 0.2. The TPC-H scale factor (SF) ranges from 1 to 32. We observe that, as the data size increases, the runtime of the \textit{All-Pairs} method increases quadratically, while the runtime of the \textit{on-the-fly Index} method has a linear speedup. Moreover, the speedup results in the figure demonstrate that the \textit{on-the-fly Index} method is approximately three orders of magnitude faster than \textit{All-Pairs SGB-Any} as the data size increases.
\end{sloppypar}

\subsection{Runtime Comparison with Clustering Algorithms}

\begin{figure}
        \centering
        \begin{subfigure}[b]{0.23\textwidth}
                \includegraphics[width=\columnwidth]{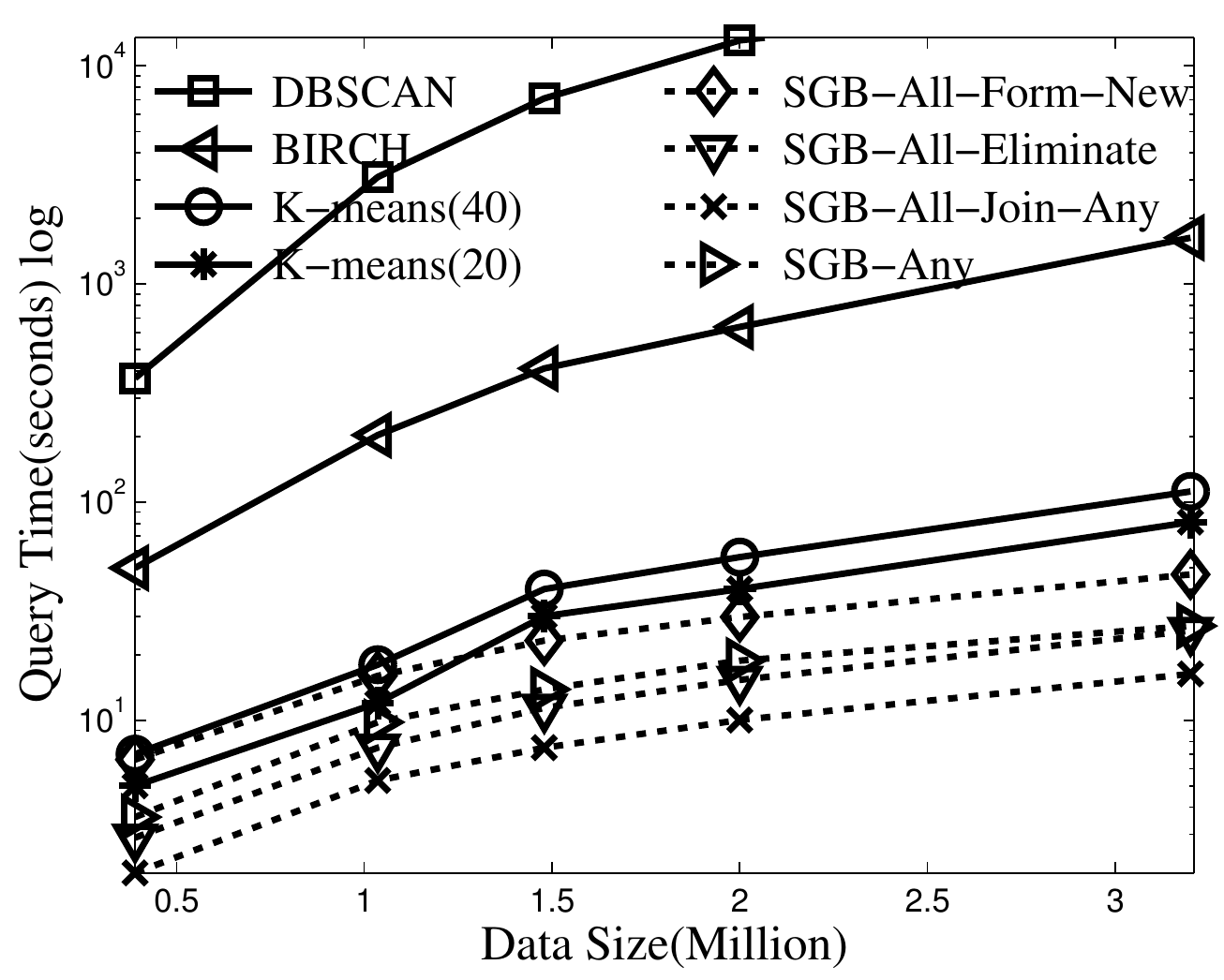}
                \caption{Runtime on Brightkite}
                \label{fig:Brightkite}
        \end{subfigure}%
        ~ 
        \begin{subfigure}[b]{0.23\textwidth}
                \includegraphics[width=\columnwidth]{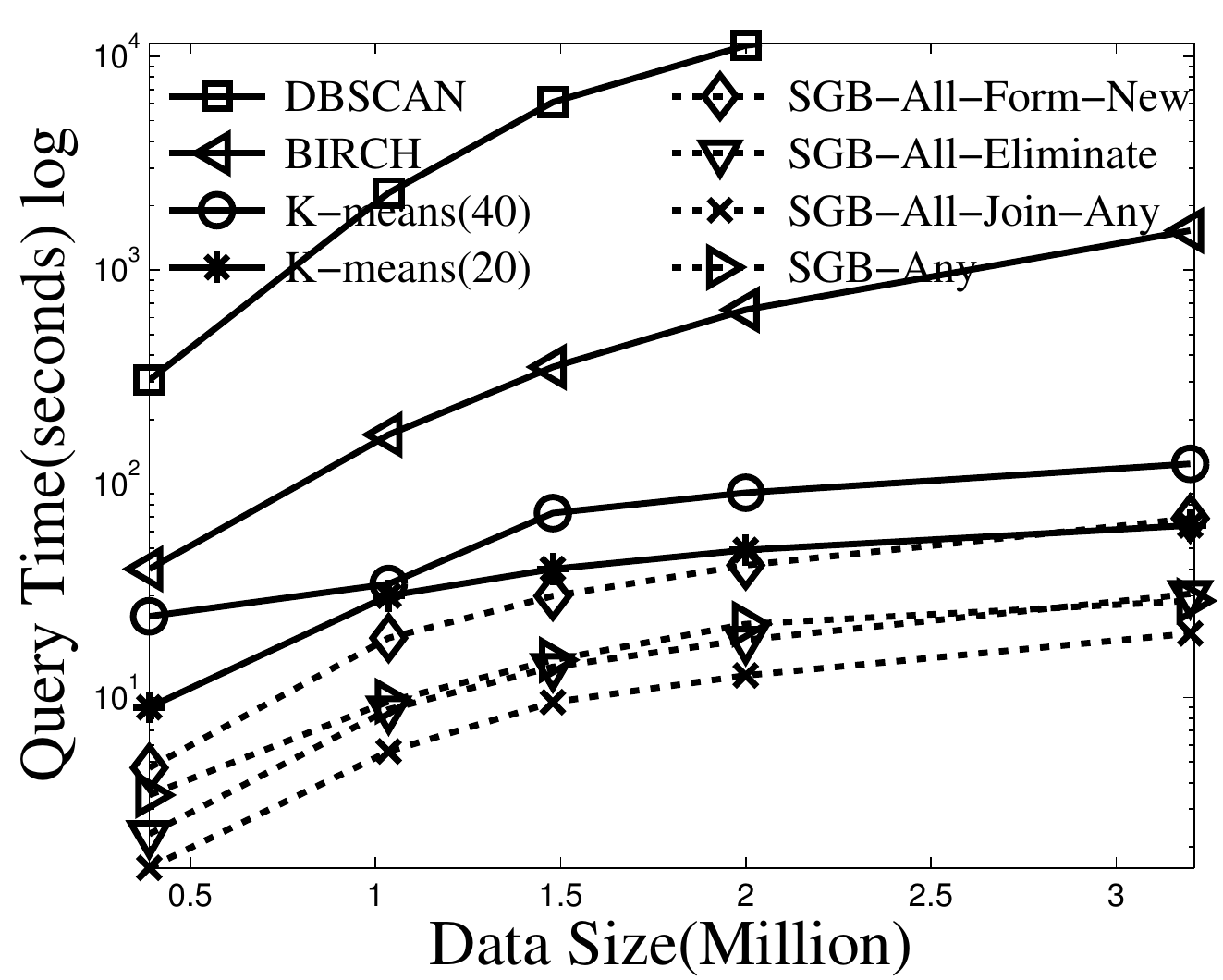}
                \caption{Runtime on Gowalla}
                \label{fig:Gowalla}
        \end{subfigure}
        \caption{SGB vs Clustering Algorithm}
        \label{fig:SGB_clustering}
\end{figure}

We compared the runtime of our SGB operators with three clustering algorithms, namely,
\textit{K-means}~\cite{BIBExample:kanungo2002efficient}, \textit{DBSCAN}~\cite{BIBExample:zhang1996birch}, and \textit{BIRCH}~\cite{BIBExample:ester1996density}. 
Specifically,  we use the state-of-the-art implementation of \textit{DBSCAN} with an R-tree 
from~\cite{Achtert2013v6}, 
the similarity threshold $\epsilon$ for both \textit{DBSCAN} and SGB is set i $0.2$, and 
the parameter \textit{K} of \textit{K-means} is set to $20$ and $40$, respectively.  
Figure~\ref{fig:SGB_clustering} shows the proposed SGB operations significantly outperform \textit{DBSCAN}, \textit{BIRCH} and \textit{K-means} by 1 to 3 order of magnitude on the real-world data respectively. The main reason is that the clustering algorithms
scan the data more than once for convergence. On the contrary, SGB operations compute groups on-the-fly, and use group bounda and a spatial index to reduce the overhead of distance computation with processed tuples. In addition, clustering algorithms have to read data from the database system making them slower than our built-in SGB operations.

\subsection{Overhead of SGB}
\begin{figure}
        \centering
        \begin{subfigure}[b]{0.23\textwidth}
                \includegraphics[width=\columnwidth]{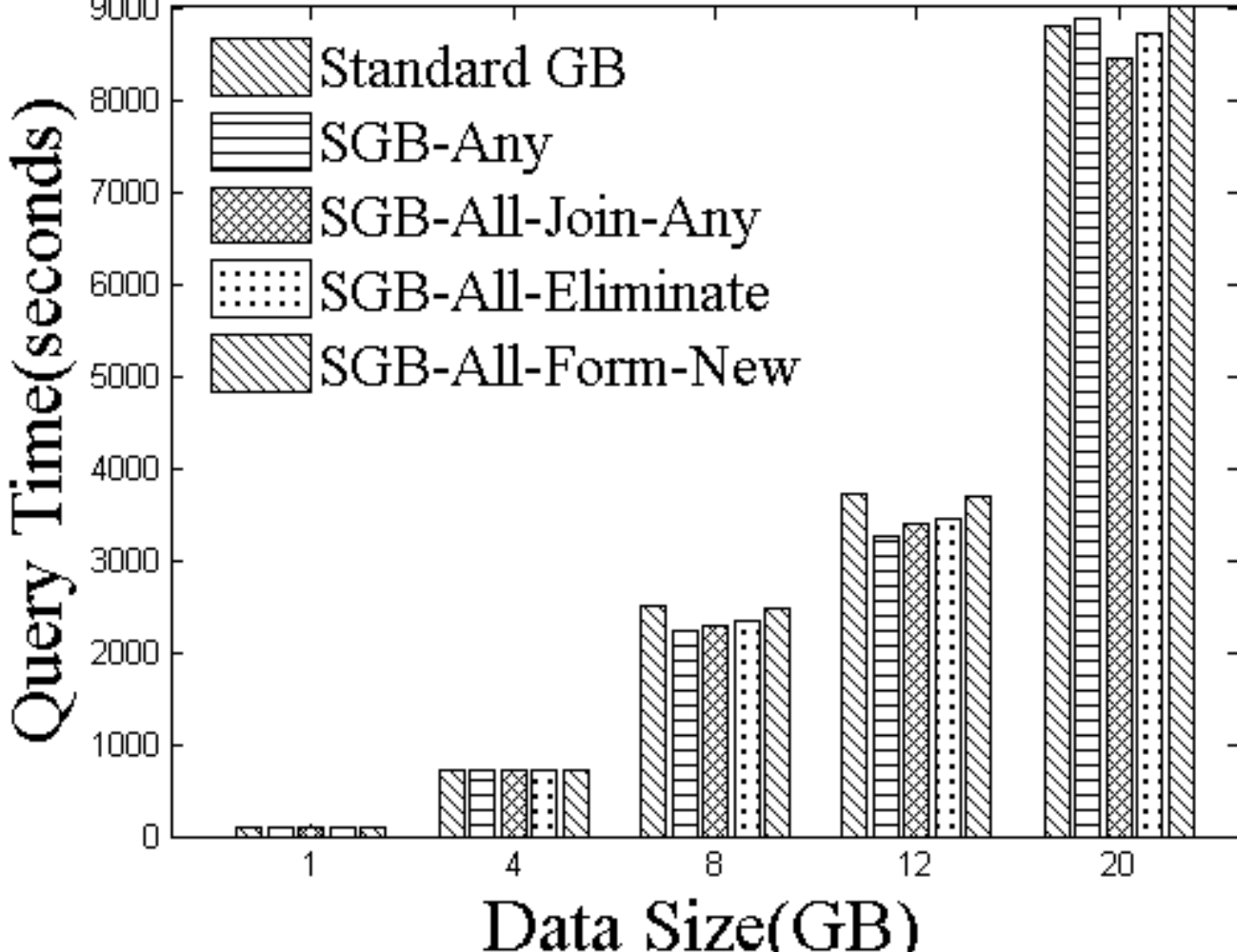}
                \caption{ GBY2 vs SGB3 and SGB4}
                \label{fig:GBY234}
        \end{subfigure}%
        ~ 
        \begin{subfigure}[b]{0.23\textwidth}
                \includegraphics[width=\columnwidth]{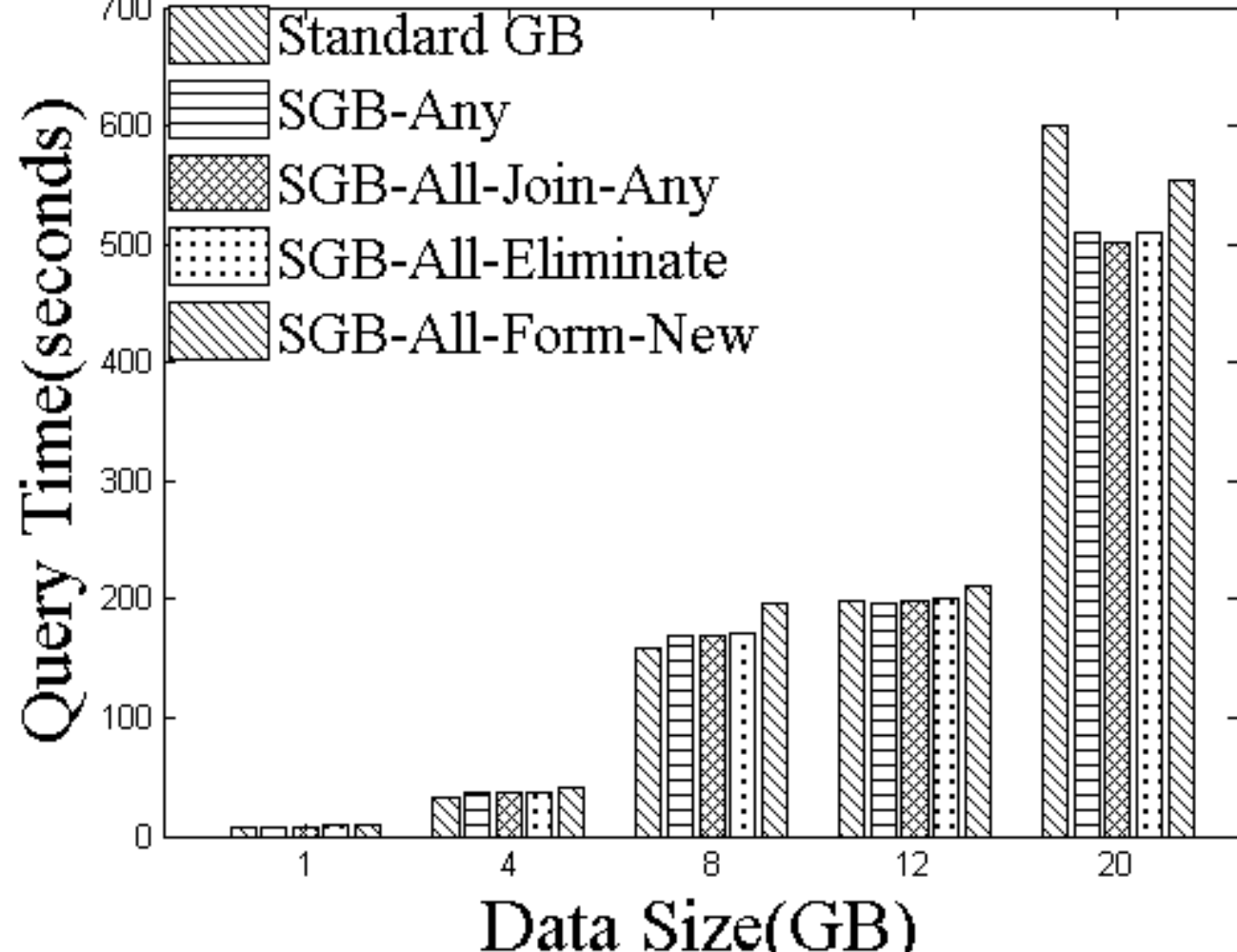}
                \caption{ GBY3 vs SGB5 and SGB6}
                \label{fig:GBY356}
        \end{subfigure}
        \caption{The effect of the data size on SGB vs. SQL GBY.}
        \label{fig:SGB_overhead}
\end{figure}

Figure~\ref{fig:SGB_overhead} illustrates the effect of the various data sizes on the runtime of similarity-based groupings and traditional Group-By queries while varying the scale factor from 1G to 20G.
The similarity threshold $\epsilon$ is fixed to 0.2. The semantics of the ON-OVERLAP clause plays a key role on the runtime of SGB-All. For instance, the JOIN-ANY variant achieves the best runtime among the SGB-All variants as it places overlapped elements into arbitrarily chosen groups. On the contrary, the FORM-NEW-GROUP incurs additional runtime cost while placing overlapped elements into new groups. The ELIMINATE semantics drops all overlapped elements causing the size of the output groups to shrink. Furthermore, the performance of traditional Group-by operator is comparable to the SGB-All and SGB-Any variants when using the \textit{on-the-fly Index}. For instance, The SGB-All ON-OVERLAP JOIN-ANY shows better performance than that of traditional Group-By. The SGB-All ON-OVERLAP ELIMINATE, SGB-All ON-OVERLAP FORM-NEW and SGB-Any shows 15 percent, 40 percent and 20 percent overhead than the traditional Group-By, respectively.

\section{Conclusion}
\label{section:conclusion}

In this paper, we address the problem of similarity-based grouping over multi-dimensional data. We define new similarity grouping operators with a variety of practical and useful semantics to handle overlap. We provide an extensible algorithmic framework to efficiently implement these operators inside a relational database management system under a variety of semantic flavors. The performance of SGB-All performs up to three orders of magnitude better than the naive \textit{All-Pairs} grouping method. Moreover, the performance of the optimized SGB-Any performs more than three orders of magnitude better than the naive approach. Finally, the performance of the proposed SGB operators is comparable to that of standard relational Group-by.

{
\scriptsize
\bibliographystyle{IEEEtran}
\bibliography{sigproc}

\begin{thebibliography}{10}
\providecommand{\url}[1]{#1}
\csname url@samestyle\endcsname
\providecommand{\newblock}{\relax}
\providecommand{\bibinfo}[2]{#2}
\providecommand{\BIBentrySTDinterwordspacing}{\spaceskip=0pt\relax}
\providecommand{\BIBentryALTinterwordstretchfactor}{4}
\providecommand{\BIBentryALTinterwordspacing}{\spaceskip=\fontdimen2\font plus
\BIBentryALTinterwordstretchfactor\fontdimen3\font minus
  \fontdimen4\font\relax}
\providecommand{\BIBforeignlanguage}[2]{{%
\expandafter\ifx\csname l@#1\endcsname\relax
\typeout{** WARNING: IEEEtran.bst: No hyphenation pattern has been}%
\typeout{** loaded for the language `#1'. Using the pattern for}%
\typeout{** the default language instead.}%
\else
\language=\csname l@#1\endcsname
\fi
#2}}
\providecommand{\BIBdecl}{\relax}
\BIBdecl

\bibitem{BIBExample:han2006data}
J.~Han, M.~Kamber, and J.~Pei, \emph{Data mining: concepts and
  techniques}.\hskip 1em plus 0.5em minus 0.4em\relax Morgan kaufmann, 2006.

\bibitem{BIBExample:silva2013similarity}
Y.~N. Silva, W.~G. Aref, P.-A. Larson, S.~S. Pearson, and M.~H. Ali,
  ``Similarity queries: their conceptual evaluation, transformations, and
  processing,'' \emph{The VLDB Journal}, vol.~22, no.~3, pp. 395--420, 2013.

\bibitem{BIBExample:adali1998multi}
S.~Adali, P.~Bonatti, M.~L. Sapino, and V.~Subrahmanian, ``A multi-similarity
  algebra,'' in \emph{ACM Sigmod Record}, vol.~27, no.~2.\hskip 1em plus 0.5em
  minus 0.4em\relax ACM, 1998, pp. 402--413.

\bibitem{BIBExample:atnafu2001similarity}
S.~Atnafu, L.~Brunie, and H.~Kosch, ``Similarity-based operators and query
  optimization for multimedia database systems,'' in \emph{IDEAS}, 2001, pp.
  346--355.

\bibitem{BIBExample:braunmuller2001multiple}
B.~Braunmuller, M.~Ester, H.-P. Kriegel, and J.~Sander, ``Multiple similarity
  queries: A basic dbms operation for mining in metric databases,'' \emph{KDE},
  vol.~13, no.~1, pp. 79--95, 2001.

\bibitem{BIBExample:chen2003similar_join}
J.~Y. Chen and J.~V. Carlis, ``Similar\_join: extending dbms with a
  bio-specific operator,'' in \emph{SAC}, 2003, pp. 109--114.

\bibitem{BIBExample:razente2008aggregate}
H.~L. Razente, M.~C.~N. Barioni, A.~J. Traina, and C.~Traina~Jr, ``Aggregate
  similarity queries in relevance feedback methods for content-based image
  retrieval,'' in \emph{SAC}, 2008, pp. 869--874.

\bibitem{BIBExample:berkhin2006survey}
P.~Berkhin, ``A survey of clustering data mining techniques,'' in
  \emph{Grouping multidimensional data}.\hskip 1em plus 0.5em minus 0.4em\relax
  Springer, 2006, pp. 25--71.

\bibitem{BIBExample:kanungo2002efficient}
T.~Kanungo, D.~M. Mount, N.~S. Netanyahu, C.~D. Piatko, R.~Silverman, and A.~Y.
  Wu, ``An efficient k-means clustering algorithm: Analysis and
  implementation,'' \emph{PAMI}, vol.~24, no.~7, pp. 881--892, 2002.

\bibitem{BIBExample:zhang1996birch}
T.~Zhang, R.~Ramakrishnan, and M.~Livny, ``Birch: an efficient data clustering
  method for very large databases,'' in \emph{ACM SIGMOD Record}, vol.~25,
  no.~2, 1996, pp. 103--114.

\bibitem{BIBExample:guha1998cure}
S.~Guha, R.~Rastogi, and K.~Shim, ``Cure: an efficient clustering algorithm for
  large databases,'' in \emph{SIGMOD}, vol.~27, no.~2.\hskip 1em plus 0.5em
  minus 0.4em\relax ACM, 1998, pp. 73--84.

\bibitem{BIBExample:ester1996density}
M.~Ester, H.-P. Kriegel, J.~Sander, and X.~Xu, ``A density-based algorithm for
  discovering clusters in large spatial databases with noise,'' in \emph{KDD},
  vol.~96, 1996, pp. 226--231.

\bibitem{BIBExample:schallehn2004efficient}
E.~Schallehn, K.-U. Sattler, and G.~Saake, ``Efficient similarity-based
  operations for data integration,'' \emph{Data \& Knowledge Engineering},
  vol.~48, no.~3, pp. 361--387, 2004.

\bibitem{BIBExample:zhang2007cluster}
C.~Zhang and Y.~Huang, ``Cluster by: a new sql extension for spatial data
  aggregation,'' in \emph{GIS}, 2007, p.~53.

\bibitem{BIBExample:razente2006siren}
M.~C.~N. Barioni, H.~Razente, A.~Traina, and C.~Traina~Jr, ``Siren: A
  similarity retrieval engine for complex data,'' \emph{VLDB}, pp. 1155--1158,
  2006.

\bibitem{BIBExample:guliato2009postgresql}
D.~Guliato, E.~V. de~Melo, R.~M. Rangayyan, and R.~C. Soares, ``Postgresql-ie:
  An image-handling extension for postgresql,'' \emph{Journal of digital
  imaging}, vol.~22, no.~2, pp. 149--165, 2009.

\bibitem{BIBExample:guttman1984r}
A.~Guttman, \emph{R-trees: A dynamic index structure for spatial
  searching}.\hskip 1em plus 0.5em minus 0.4em\relax ACM, 1984, vol.~14, no.~2.

\bibitem{BIBExample:de2008computational}
M.~De~Berg, O.~Cheong, M.~van Kreveld, and M.~Overmars, \emph{Computational
  geometry}.\hskip 1em plus 0.5em minus 0.4em\relax Springer, 2008.

\bibitem{BIBExample:Tarjan}
\BIBentryALTinterwordspacing
R.~E. Tarjan and J.~van Leeuwen, ``Worst-case analysis of set union
  algorithms,'' \emph{J. ACM}, vol.~31, no.~2, pp. 245--281, Mar. 1984.
  [Online]. Available: \url{http://doi.acm.org/10.1145/62.2160}
\BIBentrySTDinterwordspacing

\bibitem{BIBExample:softonline}
\BIBentryALTinterwordspacing
``Tpc-h version 2.15.0.'' [Online]. Available: \url{http://www.tpc.org/tpch/}
\BIBentrySTDinterwordspacing

\bibitem{standford_socialdata_2011}
E.~Cho, S.~A. Myers, and J.~Leskovec, ``Friendship and mobility: User movement
  in location-based social networks,'' in \emph{Proceedings of the 17th ACM
  SIGKDD}.\hskip 1em plus 0.5em minus 0.4em\relax ACM, 2011, pp. 1082--1090.

\bibitem{Achtert2013v6}
\BIBentryALTinterwordspacing
E.~Achtert, H.-P. Kriegel, E.~Schubert, and A.~Zimek, ``Interactive data mining
  with 3d-parallel-coordinate-trees,'' in \emph{Proceedings of the 2013 ACM
  SIGMOD International Conference on Management of Data}, ser. SIGMOD
  '13.\hskip 1em plus 0.5em minus 0.4em\relax New York, NY, USA: ACM, 2013, pp.
  1009--1012. [Online]. Available:
  \url{http://doi.acm.org/10.1145/2463676.2463696}
\BIBentrySTDinterwordspacing

\bibitem{BIBExample:Atallah86}
M.~J. Atallah, ``Computing the convex hull of line intersections,'' \emph{J.
  Algorithms}, vol.~7, no.~2, pp. 285--288, 1986.

\end{thebibliography}
}

\begin{appendix}

 We analyze the runtime of SGB-All and SGB-Any. Let $n$, $k$, $|G|$, $|G_c|$, $|G_v|$ be the data cardinality, the expected number of points per group, the number of existing groups, the size $Candidate\-Groups$, and the size of $Overlap\-Groups$, respectively, where $k\leq n$ and $|G|\leq n$ as each point can belong to only one group.
\renewcommand{\thesubsection}{\Alph{subsection}}

\subsection{SGB-All}

\begin{sloppypar}
The runtime for SGB-All is output-sensitive and is influenced by several factors e.g., the ON-OVERLAP options, and the runtimes of $Find\-Close\-Groups$ and $Process\-Over\-lap$. These factors vary with $\epsilon$ and with the data distribution. For instance, the number of Groups $|G|$ can vary from 1 to $n$ depending on the value of $\epsilon$. For example, when $\epsilon$ is very small, $|G|=n$. Next, we analyze the runtime complexity for \textit{Bounds-Checking} and, the \textit{on-the-fly index for Bounds-Checking} using the various ON-OVERLAP options.

\textbf{SGB-All Join-Any}.
Refer to Procedure~\ref{alg:findcloseboundchecking} \textit{Bounds-Checking}. It finds the groups $CandidateGroups$ by linearly testing all existing groups (Lines 4-6) to determine if  point $p_i$ can join Group $g_j$. Each test takes constant time.
Thus, the runtime of ON-OVERLAP JOIN-ANY is bounded by the number of groups, i.e., $O(n\:|G|)$.

Refer to Procedure~\ref{alg:findclose-boundindex}. $Groups\_IX$ is an \textit{on-the-fly}  R-tree that indexes the bounding rectangles of all existing groups. Given a new data point, say $p_i$, a window query of size $2\epsilon$ on $Groups\_IX$ finds the groups $Candidate\-Groups$ that $p_i$ can join. Thus, the runtime for Procedure~\ref{alg:findclose-boundindex} (Line~4) is $O(log\;|G|)$ and the overall runtime of ON-OVER\-LAP JOIN-ANY is $O(n\;log\;|G|)$. When $|G| = n$ (the number of inputs tuples), the worst-case runtime of the \textit{on-the-fly Index for Bounds-Checking} ON-OVERLAP JOIN-ANY is no better than $O(n\;log\;n)$. In contrast, when $|G|$ is constant, e.g., 1, the best-case runtime is $O(n)$. Finally, the average-case runtime of the \textit{on-the-fly index for Bounds-Checking} is $O(n\;log\;|G|)$.

\textbf{SGB-All Eliminate}. The semantics of ON-OVERLAP ELIMINATE incurs additional ($k\;|G_v|$) time while inspecting Set  $OverlapGroups$ to retrieve the subset that satisfies the similarity predicate (Lines 8-10) in Procedure~\ref{alg:findcloseboundchecking} and (Lines 10-12) in Procedure~\ref{alg:findclose-boundindex}).
In addition, $ProcessEliminate$ (Line 13) in Procedure~\ref{alg:processgroupingall} incurs additional cost of $|G_c|$ to update the bounds of the candidates groups after removing the overlapped points. Thus, the runtime of \textit{Bounds-Checking} ON-OVERLAP ELIMINATE is $O(n\;(|G| +|G_c|+ |G_v|\; k))$ while the runtime of \textit{on-the-fly Index for Bounds-Checking} ON-OVERLAP ELIMINATE is $O(n\;(log\;|G| + |G_c| + |G_v|\; k))$. Naturally, $k=n/|G|$, so the runtime of \textit{on-the-fly Index for Bounds-Checking} ON-OVERLAP ELIMINATE is $O(n\;(log\;|G| + |G_c|+n \;|G_v|/|G|))$. In the worst-case, $|G|=n$,  $|G_c|=|G|$ and $|G_v|/|G|=constant$, and the corresponding runtime of \textit{on-the-fly Index for Bounds-Checking} ON-OVERLAP ELIMINATE is $O(n^2)$. In contrast, the best-case runtime is $O(n)$ when the sizes $|G| = |G_v| = |G_c| = 1$. The average-case runtime is  $O(n\;log\;|G|)$ when the sizes of $OverlapGroups$ $|G_v| \ll n$ and $CandidateGroups$  $|G_c| \ll n$.

\textbf{SGB-All FORM-NEW-GROUP}.
Procedures~$Process\-New\-Group$ and $Process\-Overlap\-New\-Group$ insert the overlapped points into a temporary set $S'$.
Upon finding all points in $S'$, SGB-All recursively performs a new round of Form-NEW-GROUP while grouping the contents of $S'$ until $S'$ is empty.
Let $m$ be the recursion counter that is initially 0, and $S'_m$ be the set $S'$ at recursion stage m. Then, $S'_0$ is the input dataset where the size of $S'_0$ i.e., $|S'_0| = n$.
The time cost for each round is
$t_m$= O($|S'_m| \; O(Find\-Close\-Groups\-ALL)+O(Process\-Overlap)$) that is $t_m$=$O(|S'_m|\;(|G^m| + |G^m_{c}| + |G^m_{v}|\; k^m)$, where $|G^m|$, $|G^m_{c}|$ and $|G^m_{v}|$ are the number of existing groups, $Candidate\-Groups$, and $Overlap\-Groups$  at each round $m$, respectively.
Thus, the overall runtime of SGB-All FORM-NEW-GROUP is the sum of $t_m$ from recursion depth 0 to $DP$, where $t_m$ is the
cost at Recursion Depth $m$. Then,
the complexity of \textit{Bounds-Checking} is
$\sum_{m=0}^{DP} t_m$
=
$\sum_{m=0}^d O(|S'_m|\;(|G^m| + |G^m_{c}| + |G^m_{v}|\; k^m))$.
Similarly, the time complexity of the \textit{on-the-fly index} for Bounds-Checking is $\sum_{m=0}^d O(S'_m\;$ $(log\;|G^m| + |G^m_{c}| + |G^m_{v}|\;k_{m}))$. The best-case behavior of \textit{Index Bounds-Checking} for FORM-NEW-GROUP occurs when set $OverlapGroups$ is empty and the size of $CandidateGroups$ is constant. Then, the best-case runtime is $O(n)$. In contrast, if the recursion depth is almost $n$, the worst-case runtime is $O(n^3)$. On average, the recursion counter $m=constant \ll n$ and $|S'_m| \ll n$, and the complexity is $O(m \; n \; log(|G|))$.

\textbf{The Convex Hull Test} in Section~\ref{section:false-positive} forms a convex hull for each group $g_j$ to filter out the false-positive points.
The expected size of the convex hull for one group $g_j$ is $h$, where $h=log \;k$~\cite{BIBExample:Atallah86}, where $k$ is the expected number points in $g_j$. Refer to Procedure~\ref{alg:convexhulltest}. It takes $O(log\;h)$ to test if a point is inside the convex hull (Line 2). Moreover, given a point, say $p_i$, located outside the convex hull, it takes $O(log\;h)$ to obtain the farthest point from $p_i$ (Line 5). Thus, for a group of points, $g_j$, the time to test if $p_i$ can join $g_j$ is $O(log\;h+log\;h)$; that is $O(log\;log \; k)$. \textit{ConvexHullTest} is performed for each group that passes the $PointInRectangle$ test with $O(log\;k)$  cost (using $L_\infty$). Thus, the computation cost to extend Procedures~\ref{alg:findcloseboundchecking} and~\ref{alg:findclose-boundindex} with  \textit{ConvexHullTest} is $O(n \; |G| \; log\;k)$ for \textit{Bounds-Checking} and  $O(n \; log\;|G| \; log\;k)$ for the \textit{on-the-fly Index} for Bounds-Checking. Finally, the average-case runtime of the \textit{on-the-fly Index} for Bounds-Checking when using $L_2$ is $O(n \; log\;|G| \;log\;k)$. Notice that the actual running time is faster than the average-case because the convex hull test is executed only if a new point has passed the Group $g_j$'s rectangle test.
\end{sloppypar}

\subsection{SGB-Any}

Refer to Procedure \ref{alg:findcloseany}. For each new input point $p_i$, the window query returns the processed points that are within $\epsilon$ from $p_i$. Given a set of $n$ points, the complexity of the window query is $O(n\;log\;n)$. Moreover, Procedures $getGroups$ and $MergeGroupsInsert$ use Union-Find to keep track of new, existing, and merged groups. The amortized runtime of Union-Find for $n$ points is $O(m'\alpha(n))$ \cite{BIBExample:Tarjan}, where $m'$ is the total operations to build new groups, $m'=|G|$, $\alpha(n)$ is a very slowly growing function, and $\alpha(n)\leq 4$. Therefore, the average case of Union-Find running time is $O(n)$, where $m'\leq n$. Hence, the average-case runtime of SGB-Any usinpg an on-the-fly index is $O(n\;log\;n)+O(n)$, that is $O(n\; log\;n)$. Also, using $L_2$ requires an additional step $(verifyPoints)$ to filter out the points that do not satisfy the
similarity predicate in \textit{OverlapGroups} (Line 7) with a cost $k'$ per point, where $k'$ is the expected number of points within a window query. Consequently, the runtime cost of SGB-Any using $L_2$ is $O(n \; log\;n + n \; k')$. $k'$ is influenced by $\epsilon$. Thus, the worst-case runtime when using $L_2$ is $n^2$, when $k'\approx n$. If $k'$ is constant, the average-case runtime is $O(n\; log\;n)$.
The average-case runtime of the \textit{on-the-fly Index} for SGB-Any is $O(n \; log\;n)$ for both  $L_{\infty}$  and $L_2$.


\end{appendix}


\end{document}